\documentclass[12pt,journal,onecolumn]{IEEEtran}  

\usepackage{amsmath,amssymb,nicefrac,bm,upgreek,mathtools,verbatim,enumerate,cite}
\usepackage[final]{hyperref}
\usepackage[mathscr]{eucal}
\usepackage{dsfont} 
\usepackage[normalem]{ulem}
\usepackage{graphicx}

\graphicspath{{./Figures/}}
\newtheorem{definition}{Definition}[section]
\newtheorem{lemma}{Lemma}[section]
\newtheorem{theorem}{Theorem}[section]

\newtheorem{remark}{Remark}[section]

\newtheorem{assumption}{Assumption}[section]

\newtheorem{corollary}{Corollary}[section]

\allowdisplaybreaks
\hypersetup{
  colorlinks=true,
  citecolor=mblue,
  linkcolor=mblue,
  frenchlinks=false,
  pdfborder={0 0 0},
  breaklinks}
\usepackage[capitalize,nameinlink]{cleveref}

\crefname{section}{Section}{Sections}
\crefname{subsection}{subsection}{subsections}
\crefname{notation}{Notation}{Notations}
\crefname{hypothesis}{Hypothesis}{Conditions}
\crefname{assumption}{Assumption}{Assumptions}

\Crefname{figure}{Figure}{Figures}

\crefformat{equation}{\textup{#2(#1)#3}}
\crefrangeformat{equation}{\textup{#3(#1)#4--#5(#2)#6}}
\crefmultiformat{equation}{\textup{#2(#1)#3}}{ and \textup{#2(#1)#3}}
{, \textup{#2(#1)#3}}{, and \textup{#2(#1)#3}}
\crefrangemultiformat{equation}{\textup{#3(#1)#4--#5(#2)#6}}%
{ and \textup{#3(#1)#4--#5(#2)#6}}{, \textup{#3(#1)#4--#5(#2)#6}}%
{, and \textup{#3(#1)#4--#5(#2)#6}}

\Crefformat{equation}{#2Equation~\textup{(#1)}#3}
\Crefrangeformat{equation}{Equations~\textup{#3(#1)#4--#5(#2)#6}}
\Crefmultiformat{equation}{Equations~\textup{#2(#1)#3}}{ and \textup{#2(#1)#3}}
{, \textup{#2(#1)#3}}{, and \textup{#2(#1)#3}}
\Crefrangemultiformat{equation}{Equations~\textup{#3(#1)#4--#5(#2)#6}}%
{ and \textup{#3(#1)#4--#5(#2)#6}}{, \textup{#3(#1)#4--#5(#2)#6}}%
{, and \textup{#3(#1)#4--#5(#2)#6}}

\crefdefaultlabelformat{#2\textup{#1}#3}
\newcommand{\rcnt}{{\mathscr{R}_{\mathsf{net}}}}
\newcommand{\rcp}{{\mathscr{R}_{\mathsf{p}}}}
\newcommand{\rc}{{\mathscr{R}}}

\newcommand{\SSp}{\mathcal{S}}       
\newcommand{\QSp}{\mathbb{Q}}        
\newcommand{\cZ}{\mathcal{Z}}        
\newcommand{\Uadm}{\mathfrak{U}}     

\newcommand{\hUsm}{\widehat{\mathfrak{U}}_{\mathrm{sm}}}

\newcommand{\cB}{\mathcal{B}}        

\newcommand{\cN}{\mathcal{N}}        
\newcommand{\pdef}{\mathcal{M}^{+}}
\newcommand{\psdef}{\mathcal{M}^{+}_{0}}
\newcommand{\cT}{\mathcal{T}}
\DeclareMathOperator{\trace}{Tr}
\DeclareMathOperator{\dist}{dist}

\newcommand{\Lyap}{\mathscr{V}}
\newcommand{\rvi}{{\widetilde\varphi}}

\newcommand{\Rd}{{\mathbb{R}^d}}
\newcommand{\RR}{\mathbb{R}} 
\newcommand{\NN}{\mathbb{N}} 
\newcommand{\Prob}{\mathbb{P}} 
\newcommand{\Exp}{\mathbb{E}} 

\newcommand{\df}{\coloneqq} 
\newcommand{\transp}{^{\mathsf{T}}}
\newcommand{\abs}[1]{\lvert #1 \rvert} 
\newcommand{\norm}[1]{\lVert #1 \rVert} 

\newcommand{\indic}[1]{\mathds{1}_{#1}}
\newcommand*\wbar[1]{%
  \vbox{%
    \hrule height 0.5pt
    \kern0.25ex
    \hbox{%
      \kern-0.1em
      \ifmmode#1\else\ensuremath{#1}\fi
      \kern-0.1em
    }
  }
}

\DeclareMathOperator*{\Argmin}{Arg\,min}

\DeclareMathOperator*{\spn}{span}

\usepackage{color}
\definecolor{mred}{rgb}{.5,.0,.0}
\definecolor{dmagenta}{rgb}{.4,.1,.5}
\definecolor{dblue}{rgb}{.0,.0,.5}
\definecolor{mblue}{rgb}{.0,.0,.7}
\definecolor{ddblue}{rgb}{.0,.0,.4}
\definecolor{dred}{rgb}{.7,.0,.0}
\definecolor{dgreen}{rgb}{.0,.5,.0}
\definecolor{Eeom}{rgb}{.0,.0,.5}

\begin{document}

\title{Optimal Sensor Scheduling under Intermittent Observations
Subject to Network Dynamics}
\author{Hassan Hmedi$^\ddagger$, 
Johnson~Carroll$^\dagger$,~\IEEEmembership{Senior~Member,~IEEE}, and 
Ari~Arapostathis$^*$$^\ddagger$,~\IEEEmembership{Fellow,~IEEE}

\thanks{$^*$This research was supported in part by the Army Research Office
through grant W911NF-17-1-001,  and in part by Office of Naval Research
through grant N00014-16-1-2956
and was approved for public release under DCN \#43-4982-19.}
\thanks{$^\dagger$J. Carroll is with the Faculty of Engineering and the Built Environment, University of Johannesburg, Johannesburg,
2006, South Africa (e-mail: jcarroll@uj.ac.za).}
\thanks{$^\ddagger$
H. Hmedi and A. Arapostathis are with the Electrical and Computer
Engineering Department, 2501 Speedway, EER 7.824,
The University of Texas at Austin, Austin, TX 78712
(e-mail: $\lbrace$hmedi,ari$\rbrace$@utexas.edu).}}

\markboth{Optimal Sensor Scheduling under Intermittent Observations}%
{J. Carroll, H. Hmedi, and A. Arapostathis}

\maketitle
\thispagestyle{empty}

\begin{abstract}
Motivated by various distributed control applications,
we consider a linear system with Gaussian noise 
observed by multiple sensors which transmit measurements 
over a dynamic lossy network. 
We characterize the stationary optimal sensor scheduling policy 
for the finite horizon, discounted, and long-term average cost problems 
and show that the value iteration algorithm converges to a 
solution of the average cost problem. 
We further show that the suboptimal policies provided
by the rolling horizon truncation of the value iteration
also guarantee stability and provide
near-optimal average cost. 
Lastly, we provide qualitative characterizations of the multidimensional
set of measurement loss rates for which the system is stabilizable
for a static network,
thus extending earlier results on intermittent observations.
\end{abstract}


\IEEEpeerreviewmaketitle
\section{Introduction}
\IEEEPARstart{M}{odern} networking structures and applications
have inspired considerable interest in remote sensing and control.
The performance of a distributed control system is highly dependent
on the structure of the communication links between system components,
and the added complexity introduced by these links makes reliably 
predicting and controlling behavior difficult. 
Further, distributed systems with multiple sensors require control 
of both the system as well as the scheduling of observations.
As this work addresses a system with both intermittent observations
and multiple sensors, we present a brief summary of related work in each
area before discussing the few efforts to 
address both aspects.

\subsubsection*{Intermittent Measurements}
A fundamental problem with distributed sensing is accounting for the possibility 
of lost or intermittent measurements, which can occur randomly
or as a result of interference (such as packet collisions). 
In the seminal work of \cite{Sinopoli2004}, it was shown that for a 
discrete time linear system with appropriate Gaussian noise, 
the error covariance is bounded provided the measurement loss rate
is below a particular critical value. 
A number of additional studies have sought to further characterize 
the behavior of the error covariance for particular systems
\cite{Mo2011b},
or with additional assumptions
\cite{Plarre2009,
Kar2012,
Mo2012}.

More complex network behavior has also been incorporated into the estimation
problem.
A number of studies have considered correlated measurement losses, 
where consecutive transmissions are more or less likely to be lost,
frequently modeled as Markov processes of varying complexity
\cite{Huang2007,
Xie2008,
Xiao2009,
Censi2011,
Rohr2014}.
Some have additionally incorporated measurement transmission delays
\cite{Xiong2007,Wu2015}
and additional network constraints 
such as data transmission rates or power considerations
\cite{You2010, 
You2011,You2011a}.
When multiple distributed sensors broadcast on correlated channels,
as in wireless networks, 
additional intermittency issues arise from interference 
\cite{Gupta2007,
Zhang2011,
Quevedo2013,
Sui2015}.

Previous studies have also a variety of transmission contexts.
When the sensor has local processing capability, 
a local state estimate can be calculated and transmitted in lieu of the 
measurement \cite{Shi2010,
Mo2012a}.
Other studies have considered networks in which successful data transmission
is not consistently verifiable or acknowledged \cite{Sinopoli2008,Garone2009},
or data transmissions may be partially lost \cite{Wang2009},
and some have sought to extend the distributed control problem to 
include intermittent actuation
\cite{Garone2012,
Sun2014,
Yu2015}
and intermittent acknowledgement messages 
\cite{Moayedi2013}.

\subsubsection*{Sensor Scheduling}
An aspect of distributed sensing considers situations in which 
network characteristics constrain transmission, forcing the
system to choose if a sensor will transmit at each time step. 
Sensor schedules attempt to maintain system stability while 
optimizing system performance according to various metrics. 
Some approaches scheduled sensor transmissions randomly according 
to a predetermined (possibly random) schedule
\cite{Gupta2006,
Battistelli2012,
Mo2011}.

Dynamic sensor scheduling, based on various information available to the scheduler, 
can lead to significantly better performance but requires more complex analysis.
Many studies trigger transmission when a sensor's information will substantially
improve the state estimate of the system
\cite{Trimpe2014,
Han2015,
Weerakkody2016},
while more complex schemes seek the best global sensor transmission policy 
based on network structure and constraints. 

In \cite{Wu2008},
the authors characterize the optimal sensor schedule for the LQG control problem, 
and show that a partial separation principle allows a full characterization
of the finite horizon and long-term average control problems. 
The resulting dynamic programming can be computationally difficult 
even for the finite horizon problem, so
other approaches have included relaxation to allow convex optimization 
\cite{Joshi2009,
Mo2011a},
eliminating redundant schedules \cite{Vitus2012},
or finding simpler, near-optimal periodic policies
\cite{Zhao2014}.
For the simpler case of multiple independent linear systems, 
\cite{Han2017}
showed similar results and analyzed various suboptimal policies. 

\subsubsection*{Sensor scheduling with intermittent measurements}
The intersection of these two areas, namely optimal sensor scheduling with 
intermittent network links, has been largely neglected.
An early example is \cite{Ambrosino2009},
which considers a specific network model based on packet protocols and 
selects sensors to minimize the instantaneous error covariance. 
Following the pattern of earlier work, the authors use a convex relaxation
of the optimization problem to find a suboptimal schedule. 

Another early effort considers the finite-horizon sensor scheduling problem 
when measurements are randomly lost with fixed probability \cite{Jia2012}. 
Rather than seeking optimal scheduling policies, however, the authors
present a heuristic rule
(that sensors with less measurement
noise should transmit as much as possible but later than noisier sensors)
and show that there exists a policy obeying the rule which is at least as good
as any policy that violates the rule. 

The authors in \cite{Mo2014} present a more comprehensive linear estimation problem,
proving the independence of the average cost from the initial estimation error 
(previously shown in \cite{Wu2008})
and the existence of near-optimal periodic schedules 
(simultaneously developed in \cite{Zhao2014}). 
However, \cite{Mo2014} also briefly extends these results to schedule-dependent
Bernoulli measurement losses.

More recently in \cite{Leong2017}, the authors considered optimal scheduling of sensors, 
each of which can fully observe the unstable system states and calculates a local
state estimate
for transmission to a central estimator. 
Since the local estimates converge for the infinite horizon problem, 
one need only consider countably many possible covariance values
when selecting sensors. 
The study in \cite{Leong2017} also describes several structural results for the 
optimal schedules, and goes on to mention transmitting measurements
instead of estimates. However, the structural results do not hold
for measurement transmission, and the authors balk at 
considering the entire set of positive semidefinite matrices
as a state space for the sensor scheduling problem.
Lastly, we want to mention the preprint in \cite{Arxiv-1804}, which
provides an index-bases heuristic to avoid hard computations.

\begin{figure}[!t]
\centering
\includegraphics[width=0.5\columnwidth]{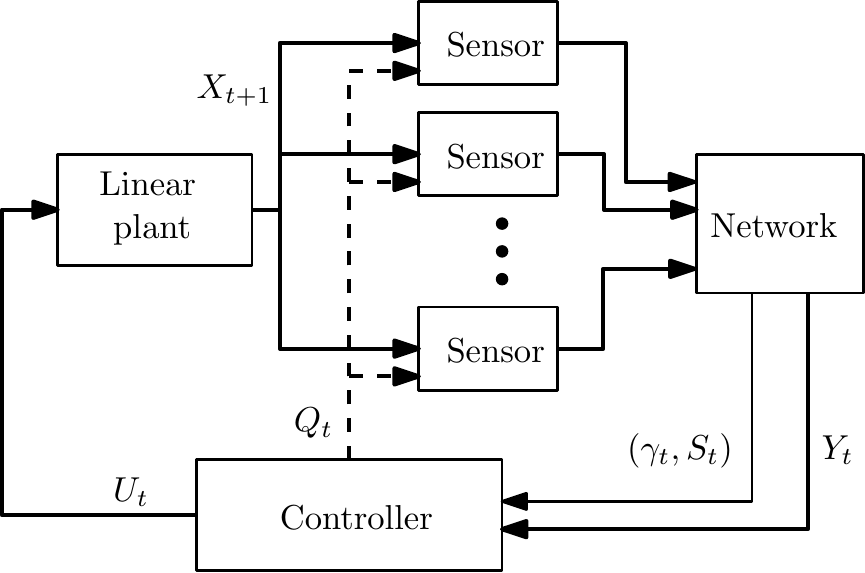}
\caption{Overview of the system detailed in Section \ref{sec-system}.
An observation $Y_{t}$ is lost ($\gamma_{t}=0$) 
with probability $\lambda_{t}$, which depends on which
sensor is queried ($Q_{t}$) and network state $(S_{t})$.}
\label{fig:SystemDiagram}
\end{figure}

In this work, we consider a discrete-time linear quadratic Gaussian (LQG) system 
observed by a finite number of sensors. 
When queried, a sensor attempts to transmit the measurement to the 
controller over a noisy network which intermittently loses the measurement.
Further, the network has its own query-dependent stochastic dynamics, allowing
for complex congestion models. 
A diagram of the system is shown in Figure \ref{fig:SystemDiagram}.

Here, we make only mild assumptions on the system structure and 
assume that the system is stabilizable. 
This rather basic assumption of stabilizability enables 
us to derive a wealth of interesting new results:
\begin{itemize}
\item A stationary, average-cost optimal policy exists, and under that
policy the system is geometrically stable.
Further, the value function can be effectively approximated by solving for the
optimal policy on 
a bounded subset of the state space and extending with any stable control. 
The process is again stable under the calculated policy, 
and the approximated average cost is an effective estimate of the true cost.
\item The value iteration algorithm, linking the finite horizon control
problem and the average cost ergodic control problem, converges
to the value function.
\item After finitely many steps, the sub-optimal policies calculated
via the value iteration algorithm also induce a stable system,
and the induced average cost converges geometrically to the optimal average cost.
\item Additionally, we show that a special case of our results generalizes
the original stabilizability results of \cite{Sinopoli2004} to the case of 
multiple scheduled sensors with unique loss rates. 
The results provide insight into the structure of a closed set
of loss rates for which the system is stabilizable and imply
that unknown loss rates can be estimated online without 
affecting long-term average performance.
We also present an explicit characterization of the stabilizable region
for independent one-dimensional systems. 
\end{itemize}

It is remarkable that such a comprehensive set of results can be shown
with only basic assumptions for a practical 
Markov decision process with an infinite state space. 
Combining key results, one can find a near-optimal, stable policy in just a few
steps of the value iteration algorithm using a truncated state space.

A version of the results in this paper appeared in \cite{CHA-CDC}.
But there are considerable differences. First, the main set of hypotheses
in \cref{A3.1} is somewhat weaker in this paper, and \cite{CHA-CDC}
does not include the technical proofs.
A major technical difference can be found in \cref{L3.2} which is crucial for the
vanishing discount method.  This is not stated correctly in \cite{CHA-CDC},
and the proof given in this paper underscores the delicate nature of this result.
This affects the results on the Bellman equation,
which are stated in stronger form here and include uniqueness of
solutions (see \cref{T3.2,T3.3}).
Lastly, the results on the relative value iteration in \cref{S4} are
quite different.

Section~\ref{sec-system} describes the system structure, our key assumptions, 
and some basic results on the Kalman filtering part of the problem.
The optimal control problems and results are presented in Section~\ref{S3},
and Section~\ref{S4} contains the results on the convergence of the 
value iteration algorithm. 
The important special case is discussed in Section~\ref{S5}, 
and finally conclusions are summarized in Section~\ref{sec-conclusion}.
For clarity, all proofs and supporting lemmas are left to appendices.

\subsection{Notation}

The letter $d$ refers to the dimension of the state space.
We let $\psdef$ ($\pdef$) denote the cone of real
symmetric, positive semi-definite (positive definite) $d\times d$ matrices. 
The identity matrix of size $d$ is denoted by $I_d$.
For a matrix $G\in\pdef$, $\underline{\sigma}(G)$ and $\overline{\sigma}(G)$
denote the smallest and largest eigenvalues of $G$, respectively.
Recall that the trace of a matrix, denoted by $\trace(\cdot)$,
acts as a norm on $\psdef$.
For $\Sigma_{1},\Sigma_{2}\in\RR^{d\times d}$, we write
$\Sigma_{1}\preceq\Sigma_{2}$ when 
$\Sigma_{2}-\Sigma_{1}\in\psdef$ or $\Sigma_{1} \prec \Sigma_{2}$ 
when $\Sigma_{2}-\Sigma_{1}\in\pdef$.
For two real vectors $\lambda,\phi$ indexed by some set $I$, 
we say $\lambda\le\phi$ or $\lambda< \phi$ if
for each $i\in I$, 
$\lambda_{i}\le\phi_{i}$ or $\lambda_{i}< \phi_{i}$, respectively.
A function $f\colon\psdef\to\RR$ is concave if for 
any $\Sigma_{1},\Sigma_{2}\in\psdef$, 
\begin{equation}\label{E-NotA}
f\bigl((1-\beta)\Sigma_{1}+\beta\Sigma_{2}\bigr)\,\ge\,
(1-\beta)f(\Sigma_{1})+\beta f(\Sigma_{2})
\end{equation}
for all $\beta\in [0,1]$.
Concavity for functions $f\colon\psdef\to\psdef$ is defined in
the same way,
but replacing the inequality in \cref{E-NotA} with 
the ordering $\succeq$.
We also denote a normal distribution with mean $x$ and covariance matrix
$\Sigma$ as $\cN(x,\Sigma)$. 
Given a strictly positive real function $f$ on $\SSp\times\psdef$,
where $\SSp$ is a finite set,
the $f$-norm of a function $g\colon\SSp\times\psdef\to\RR$
is given by
\begin{equation*}
\norm{g}_{f} \,\df\, 
\sup_{(s,\Sigma)\in\,\SSp\times\psdef}\frac{\abs{g(s,\Sigma)}}{f(s,\Sigma)}\,.
\end{equation*}
We denote by $\mathcal{O}(f)$ the set of real-valued functions 
on $\SSp\times\psdef$ which have finite $f$-norm
and are continuous, concave, and non-decreasing in the second argument.
We use `lsc' as an abbreviation of lower semicontinuous.

\section{System, Sensor, and Network Model}\label{sec-system}
We consider a linear quadratic Gaussian (LQG) system 
\begin{equation}\label{E-LQG}
\begin{split}
X_{t+1}&\,=\,A X_{t}+B U_{t}+D W_{t}\,,\quad t \,\ge\,0\,, \\ 
X_0&\,\sim\,\cN(x_0,\Sigma_0)\,,
\end{split}
\end{equation}
where $X_{t}\in\Rd$ is the system state, $U_{t}\in\RR^{d_u}$ is the control, 
and $W_{t}\in\RR^{d_w}$ is a white noise process. 
We assume that each ${W_t\sim\cN(0,I_{d_w})}$ is 
i.i.d. and independent of $X_0$, and that $(A,B)$ is stabilizable. 
The system is observed via a finite number of sensors scheduled or queried 
by the controller at each time step.
The query process $\{Q_{t}\}$ takes values in 
the finite set of allowable sensor queries denoted by $\QSp$.
The network congestion is modeled as a Markov chain $\{S_{t}\}_{t\in\NN_0}$, 
also controlled by $Q_{t}$,
taking values on a finite set $\SSp$ of network states: 
\begin{equation}\label{E-net}
\Prob\bigl(S_{t+1}=s' \mid S_{t}=s, Q_{t}=q\bigr) \,=\, p_{q}(s,s')\,,
\end{equation}
for $s,s'\in\SSp$, $t\ge 0$, and a known initial state ${S_0\in\SSp}$.
Without loss of generality we let $\SSp=\{1,2,\dotsc,N\}$.

A scheduled sensor attempts to send information to the controller through the
network; depending on the state of the network, the information may be received or lost.
Let $\{\gamma_{t}\}$ be a Bernoulli process indicating
if the data is lost in the network:
each observation is either received ($\gamma_{t}=1$) 
or lost ($\gamma_{t}=0$) with a probability that depends on
the network state and the query, i.e.,
\begin{equation}\label{E-loss}
\Prob(\gamma_{t+1}=0) \,=\, \lambda(S_{t},Q_{t})\,,
\end{equation}
where the loss rate $\lambda\colon\SSp\times\QSp\to [0,1)$.
The network state $S_{t}$ and the value of $\gamma_t$ are assumed to be known
to the controller at every time step.

This behavior is modeled as
\begin{align}\label{E-LQGY}
Y_{t} &\,=\, C_{Q_{t-1},S_{t-1}} X_{t}
+F_{Q_{t-1},S_{t-1}} W_{t}\,,\quad t \ge 1,
\end{align}
if $\gamma_t=1$, otherwise no observation is received.
The dimension of $Y_t$ may be variable, and naturally equals
the number of rows of $C_q$ for $q=Q_{t-1}$.
 
For each query $q\in\QSp$, 
we assume that $\det(F_{q}F_{q}\transp)\neq 0$ and 
(primarily to simplify the analysis) that $DF_{q}\transp=0$. 
Also without loss of generality, we assume that $\text{rank}(B)=N_{u}$; if not,
we restrict control actions to the row space of $B$.

The running cost $\rc$ is the sum of a positive network cost
$\rcnt\colon\SSp\times\QSp\to\RR$ 
and a quadratic plant cost $\rcp\colon\Rd\times\RR^{d_u}\to\RR$ given by
\begin{equation}\label{E-rcp}
\rcp(x,u)= x\transp R x + u\transp M u\,,
\end{equation}
where $R,M\in\pdef$. 
To help with later analysis, we choose some distinguished network state
$s_\circ\in\SSp$, which satisfies
\begin{equation*}
s_\circ\in\Argmin_{s\in\SSp}\,\Bigl(\min_{q\in\QSp}\, \rcnt(s,q)\Bigr)\,,
\end{equation*} 
and without loss of generality assume that $\rcnt(s_\circ,q) \ge 0$.

At each time $t$, the controller takes an action ${v_{t}=(Q_{t},U_{t})}$,
the system state evolves as in \cref{E-LQG}, and the network state
transitions according to \cref{E-net}. Then the observation at $t+1$ 
is either lost or received, determined by \cref{E-LQGY} and \cref{E-loss}.
The decision $v_{t}$ is non-anticipative, i.e., should depend only on the 
history $\mathcal{F}_{t}$ of observations up to time $t$ defined by
\begin{equation*}
\mathcal{F}_{t} \,\df\, \sigma(x_0,S_0,\Sigma_0,\gamma_0,S_{1},Y_{1},\gamma_{1},
\dotsc,S_{t},Y_{t},\gamma_{t}).
\end{equation*} 
Such a sequence of decisions $v=\{v_{t}\colon t\ge 0\}$ is called a policy, 
and we denote the set of admissible policies by $\Uadm$.

For an initial condition $(s_0,X_0)$ and a policy $v\in\Uadm$, let
$\Prob^{v}$ be the unique probability measure on the trajectory space, and
$\Exp^{v}$ the corresponding expectation operator. 
When necessary, the explicit dependence on (the law of) the initial
conditions or their parameters is indicated in a subscript,
such as $\Prob^{v}_{s_0,X_0}$ or $\Exp^{v}_{x_0,s_0,\Sigma_0}$.

\subsection{A partial separation principle}\label{S2A}
We have thus far described a system given by partially observed controlled {M}arkov 
chain, which we now convert to an equivalent completely observed model.
Standard theory from Markov decision processes (MDP)
tells us that the expected value of the state 
$\widehat{X}_{t}\df\Exp[X_{t}\,|\,\mathcal{F}_{t}]$ conditioned
on the history is a sufficient statistic for control purposes, when
the optimization criterion is a sum of running cost functions
as in \cref{E-rcp}.
Consider an optimal control problem over a finite horizon,
where the optimization objective takes the form
\begin{equation*}
J_{\alpha,n}^{Q,U}(s,x,\Sigma,\Pi_{\mathsf{fin}}) \,\df\, 
\Exp^{Q,U}_{s,x,\Sigma} \Biggl[\sum_{t=0}^{n-1}\alpha^{t}\bigl(\rcnt(S_{t},Q_{t})
+ \rcp(X_{t},U_{t})\bigr)
+ \alpha^n X_n\transp\Pi_{\mathsf{fin}}X_n\Biggr]\,, 
\end{equation*}
where $\Pi_{\mathsf{fin}}\in\psdef$ is a terminal cost.
Looking ahead to the study of the infinite horizon discounted problem,
we have included a discount factor $\alpha\in(0,1]$ in the objective.
The functional $J_{\alpha,n}^{Q,U}$ depends on the initial network state $s$,
the initial state of the plant
$X_0\sim\mathcal{N}(x,\Sigma)$, with $x\in\Rd$
and $\Sigma\in\psdef$,
and the admissible policy $\{Q_t,U_t\}_{t\ge0}$.
Since the plant is governed by the linear dynamics \cref{E-LQG},
the optimal control policy $\{U_t\}_{t\ge0}$.
takes the form of the linear feedback control
\begin{equation}\label{E-u*at}
U_{t} \,=\, u^{*,n}_{\alpha,t}(\widehat{X}_t)
\,=\, - K^{(n)}_{\alpha, t}\,\widehat{X}_{t}\,,
\end{equation}
with $\widehat{X}_t$ given in \cref{E-fltr1},
where the feedback gain $K^{(n)}_{\alpha, t}$ is determined by the backward recursion
\begin{equation}\label{E-Kf}
\begin{aligned}
K^{(n)}_{\alpha, t}&\,=\, \alpha (M + \alpha B\transp\Pi^{(n)}_{\alpha, t+1} B)^{-1}
B\transp\Pi^{(n)}_{\alpha, t+1} A\,, \\
\Pi^{(n)}_{\alpha, t} &\,=\, R + \alpha A\transp \Pi^{(n)}_{\alpha, t+1} A 
- \alpha A\transp \Pi^{(n)}_{\alpha, t+1} B K^{(n)}_{\alpha, t}\,,
\end{aligned}
\end{equation}
with $\Pi^{(n)}_{\alpha, n}=\Pi_{\mathsf{fin}}$.
Detailed derivations can be found in 
\cite[Sec. 5.2]{Bertsekas2005}.

The state estimate $\widehat{X}_{t}$
and the \emph{error covariance matrix} $\widehat\Pi_t$, $t=0,\dotsc,n$, given by
\begin{equation*}
\widehat\Pi_t  \,\df\, 
\Exp\bigl[(X_{t}-\widehat{X}_{t})(X_{t}-\widehat{X}_{t})\transp\bigr]\,,
\quad t\in\NN_0\,,
\end{equation*} 
can be
calculated via the Kalman filter
\begin{equation}\label{E-fltr1}
\widehat{X}_{t+1} \,=\, A\widehat{X}_{t}+ B U_{t} \\
+ \widehat{K}^{(n)}_{Q_{t},\gamma_{t+1}}(S_t,\widehat\Pi_t)
\bigl(Y_{t+1} - C_{Q_{t}, S_t}(A\widehat{X}_{t}+ B U_{t})\bigr)\,,
\end{equation}
with $\widehat{X}_0=X_0$.
The Kalman gain $\widehat{K}_{q,\gamma}$ is given by
\begin{align*}
\widehat{K}_{q,\gamma}(s,\Sigma)& \,\df\, \Xi(\Sigma)\gamma C_{q,s}\transp
\bigl(\gamma^{2} C_{q,s}\Xi(\Sigma)C_{q,s}\transp
+F_{q,s}F_{q,s}\transp\bigr)^{-1}\,, \\
\Xi(\Sigma)& \,\df\, DD\transp+A\Sigma A\transp
\end{align*}
for $\Sigma\in\psdef$,
and the error covariance process $\bigl\{\widehat\Pi_t\bigr\}_{t\in\NN_0}$
evolves on $\psdef$ as
\begin{equation}\label{E-fltr3}
\widehat\Pi_{t+1} \,=\, \Xi(\widehat\Pi_t) 
- \widehat{K}_{Q_{t},\gamma_{t+1}}(S_t,\widehat\Pi_t)C_{Q_{t},S_t}\Xi(\widehat\Pi_t)\,,
\qquad
\widehat\Pi_0=\Sigma_0\,.
\end{equation}
When an observation is lost ($\gamma_{t}=0$), 
the gain $\widehat{K}^{(n)}_{q,\gamma_{t}}\equiv0$
for any $q\in\QSp$, and 
the observer \cref{E-fltr1} simply evolves without any correction factor.

We let $\cZ\df\SSp\times\psdef$, and $z=(s,\Sigma)$ denote a generic element of
this set.
It follows from \cref{E-fltr3},
 that $Z_t\df(S_t,\widehat\Pi_t)$, $t\in\NN_0$, forms a completely observed
controlled {M}arkov chain with state space $\cZ$ and action space $\QSp$,
and transition probability (kernel)
$\widehat\cT_{q}$ which  is defined as follows.

\medskip
\begin{definition}
For a sensor query $q\in\QSp$, define $\cT_{q}\colon\psdef\to\psdef$ by
\begin{equation*}
\cT_{q}(\Sigma) \,\df\,
\Xi(\Sigma) - \widehat{K}_{q,1}(\Sigma)C_{q}\Xi(\Sigma)\,,
\end{equation*}
and an operator $\widehat\cT_{q}$ on functions 
$f\colon\SSp\times\psdef\to\RR$,
\begin{equation}\label{E-cT}
\widehat\cT_{q} f(s,\Sigma) \,=\, \sum_{s'\in\SSp} p_{q}(s,s') 
\Bigl(\bigl(1-\lambda(s,q)\bigr)f\bigl(s',\cT_{q}(\Sigma)\bigr)
+ \lambda(s,q)f\bigl(s',\Xi(\Sigma)\bigr)\Bigr)\,.
\end{equation}
By a slight abuse of notation, if $q\colon\cZ\to\QSp$ is a stationary Markov
policy, we use $\widehat\cT_{q}$ to denote the controlled kernel,
which in this case stands for $\widehat\cT_{q}(z) = \widehat\cT_{q(z)}(z)$,
with the latter as defined in \cref{E-cT}.
Denoting the corresponding conditional expectation as $\widehat\Exp^q_z$,
we have
\begin{equation}\label{E-hExp}
\widehat\cT_{q} f(z)\,=\,
\widehat\Exp^q_{z}\bigl[f(Z_1)\bigr]\\
\,=\, \widehat\Exp^{q}\bigl[f(Z_{t+1}) \bigm| Z_{t}=z\bigr]\,.
\end{equation}
We also denote by $\widehat\Prob^Q_z$ the probability measure on the path space
of the process under an admissible policy $\{Q_t\}_{t\in\NN_0}$.
\end{definition}

Admissible policies are defined as earlier, and denoted
as $\widehat\Uadm$.
However the history space is now
\begin{equation*}
\widehat{\mathcal{F}}_{t} \,\df\, \sigma(S_0,\widehat\Pi_0,S_{1},
\widehat\Pi_1,\gamma_{1},
\dotsc,S_{t},\widehat\Pi_t,\gamma_{t}).
\end{equation*}
We say that a policy $\{Q_t\}_{t\in\NN_0}$ is  Markov
if $Q_t = q_t(S_t,\widehat\Pi_t)$, and $U_t = u_t(\widehat{X}_t)$,
for some measurable maps $q_t\colon\cZ\to\QSp$ and $u_t\colon\Rd\to\RR^{d_u}$,
with $\widehat{X}_t$ and $\widehat\Pi_t$ given by \cref{E-fltr1,E-fltr3}, respectively.
We say that such a policy is stationary Markov if the map $q$  does not depend
on $t\in\NN_0$, and we denote the class of such policies
by $\hUsm$.

The solution of the finite horizon control problem is now clear.
Let
\begin{equation}\label{E-tPiat}
\widetilde\Pi^{(n)}_{\alpha,t} \,\df\, R - \Pi^{(n)}_{\alpha,t}
 + \alpha A\transp \Pi^{(n)}_{\alpha,t} A\,,
\end{equation}
and
\begin{equation}\label{E-tJan}
\widetilde{J}_{\alpha,n}(x,\Sigma,\Pi_{\mathsf{fin}})\,\df\,
x\transp\Pi^{(n)}_{\alpha,0} x + \trace(\Pi^{(n)}_{\alpha,0} \Sigma)+
 \sum_{t=1}^{n-1}\alpha^{t}\trace(\Pi^{(n)}_{\alpha,t} DD\transp)
 +\alpha^{n}\trace(\Pi_{\mathsf{fin}}DD\transp)\,.
\end{equation}
Mimicking the derivation in \cite{Wu2008}, we then see that
\begin{equation}\label{E-sepA}
J^{Q,u^*_\alpha}_{\alpha,n}(s,x,\Sigma,\Pi_{\mathsf{fin}})
\,=\,  \widetilde{J}_{\alpha,n}(x,\Sigma,\Pi_{\mathsf{fin}})
+ \widehat{J}^{Q}_{\alpha,n}(s,\Sigma;\{\widetilde\Pi^{(n)}_{\alpha,t}\})\,,
\end{equation}
where
\begin{equation}\label{E-hJan}
\widehat{J}^{Q}_{\alpha,n}(s,\Sigma;\{\widetilde\Pi^{(n)}_{\alpha,t}\})
\,\df\, \widehat\Exp^{Q}_{s,\Sigma}\Biggl[\sum_{t=0}^{n-1}\alpha^{t}
\bigl(\rcnt(S_t,Q_t)+\trace(\widetilde\Pi^{(n)}_{\alpha,t} \widehat\Pi_t)\bigr)\Biggr]\,,
\quad \widehat\Pi_0 = \Sigma\,.
\end{equation}
Define the running cost
\begin{equation}\label{E-caqt}
c^{\alpha,n}_{q,t}(s,\Sigma) \,\df\,
\rcnt(s,q) + \trace\bigl(\widetilde\Pi^{(n)}_{\alpha,t}\Sigma\bigr)\,,
\quad (s,\Sigma)\in\cZ\,,
\ t=0,\dotsc,n\,.
\end{equation}
Minimizing \cref{E-hJan} with respect to all admissible policies
$\widehat\Uadm$, can
be accomplished by the finite horizon dynamic programming equation:
\begin{equation}\label{ET3.1A}
f^{(n)}_{t}(s,\Sigma) \,=\, \min_{q\in\QSp}\,\bigl\{c^{\alpha,n}_{q,t}(s,\Sigma)
 + \alpha \widehat\cT_{q} f^{(n)}_{t+1}(s,\Sigma) \bigr\}\,,
\end{equation}
$(s,\Sigma)\in\cZ$,
for $t=0,\dotsc,n-1$, with $f^{(n)}_{n}=0$.
Then a selector $\{q^{*,n}_{\alpha,t}\}_{0\le t\le n-1}$ from the minimizer
of \cref{ET3.1A} is an optimal Markov control,
and $f^{(n)}_{0}(s,\Sigma)$ is the optimal value of \cref{E-hJan}.

In summary then, the Markov control
$v^{*,n}_{\alpha,n}\df\bigl(q^{*,n}_{\alpha,t},u^{*,n}_{\alpha,t}\bigr)_{0\le t\le n-1}$,
with $u^{*,n}_{\alpha}$ as in \cref{E-u*at},
is optimal for the finite horizon control problem, and the optimal value
$J_{\alpha,n}^* (x,s,\Sigma)$ is given by
\begin{equation}\label{ET3.1B}
\begin{aligned}
 J_{\alpha,n}^* (x,s,\Sigma,\Pi_{\mathsf{fin}}) \,=\,
 \widetilde{J}_{\alpha,n}(x,\Sigma,\Pi_{\mathsf{fin}})
+ f^{(n)}_0(s,\Sigma)\,.
 \end{aligned}
\end{equation}

It can be seen from this solution that there is no strict separation principle between estimation and control
for the LQG model with sensor scheduling, since
the running cost in \cref{E-caqt}, and therefore also
$f^{(n)}_0(s,\Sigma)$
and the optimal $q^*_{\alpha,t}$, depend on the
matrix $\widetilde\Pi_{\alpha,t}$ in \cref{E-tPiat}.
However, as we are going to see in the next section,
for infinite horizon control problems, the matrix
$\widetilde\Pi_{\alpha,t}$ does not depend on $t$, and therefore
enters as a constant parameter in the optimization problem.

We conclude this section with an important property
of the map $\widehat\cT_q$ in \cref{E-cT}.
 We sightly abuse the terminology by calling a function 
$f$ on $\SSp\times\psdef$ concave, continuous, or monotone if
$f(s,\cdot)$ has these properties, respectively, for all $s\in\SSp$.
Note that a function on $\psdef$, such as $\trace(\cdot)$, is independent
of the first coordinate if viewed as a function on 
$\SSp\times\psdef$, but that $\widehat\cT_{q}\trace(\cdot)$ depends implicitly on $s$. 
The following lemma follows easily from the definition of 
$\widehat\cT_{q}$ using standard results from, for example, \cite[Lemmas~1--2]{Gupta2006}.

\medskip
\begin{lemma}\label{L2.1}
The map $\widehat\cT_{q}$ in \cref{E-cT}
preserves continuity and lower semicontinuity of functions, and
preserves concavity and monotonicity for 
non-decreasing functions (with respect to $\preceq$).
\end{lemma}

\begin{IEEEproof}
Both $\Xi$ and $\cT_{q}$ are continuous by inspection, 
and so $\widehat\cT_{q}$ is a convex combination of continuous
functions. Hence $\widehat\cT_{q} f$ is continuous when $f$
is continuous. If $g$ is lower semicontinuous (lsc), there exists an
increasing sequence of continuous functions $f_{n}\to g$.
Each $\widehat\cT f_{n}$ is continuous and thus $\widehat\cT g$
is lsc.

$\Xi(\widehat\Pi)$ is linear in $\widehat\Pi$, so it is also concave and non-decreasing.
Concavity of $\cT_{q}$ is a standard result (see, e.g., \cite[Lemma 1]{Gupta2006}),
as is the fact that $\Sigma\succeq\Sigma'$ implies 
$\cT_{q}(\Sigma)\succeq\cT_{q}(\Sigma')$ (e.g, \cite[Lemma 2]{Gupta2006}).
Since $\widehat\cT_{q}f(s,\widehat\Pi)$ is a convex combination of 
$f(s',\Xi(\widehat\Pi))$ and the various 
possible $f(s',\cT_{q}(\widehat\Pi))$ functions, if
$f\colon\SSp\times\psdef\to\RR$ is concave and non-decreasing in
its second argument, so is $\widehat\cT_{q}f$.
\end{IEEEproof}

\section{Infinite horizon control problems}\label{S3}

Consider the dynamics in \cref{E-LQG,E-LQGY} with initial
condition $X_0\sim\mathcal{N}(x,\Sigma)\in\Rd\times\psdef$,
and initial network state $s\in\SSp$.
We are interested in finding admissible policies 
that minimize the long-term average (ergodic) cost, 
\begin{equation*}
J^{Q,U}(x,s,\Sigma) \,\df\, \limsup_{T\to\infty}\,\frac{1}{T}\,
\Exp^{Q,U}_{x,s,\Sigma}\Biggl[\,\sum_{t=0}^{T-1}\bigl(\rcnt(S_{t},Q_{t})
+ \rcp(X_{t},U_{t})\bigr)\Biggr] 
\end{equation*}
over all admissible policies $(Q_t,U_t)_{t\in\NN_0}\in\Uadm$.
We set $J^* \df \inf_{v\in\Uadm} J^{v}$.
As it turns out, $J^*$ is a constant independent of the initial
state $(x,s,\Sigma)$.
An admissible policy $(Q_t,U_t)_{t\in\NN_0}$ which attains the
infimum, that is, $J^{Q,U}= J^*$ is called optimal.

In approaching the average cost problem, we also consider
the $\alpha$-discounted cost,
\begin{equation*}
J_{\alpha}^{Q,U}(x,s,\Sigma) \,\df\, 
\Exp^{Q,U}_{x,s,\Sigma} \Biggl[\sum_{t=0}^{\infty}\alpha^{t}\bigl(\rcnt(S_{t},Q_{t})
+ \rcp(X_{t},U_{t})\bigr)\Biggr]\,,
\end{equation*}
and analogously define $J_{\alpha}^*$, and optimality of
a policy.

\subsection{The \texorpdfstring{$\alpha$}{}-discounted control problem}

It is well known that
in order for the average cost of the plant,
with a quadratic running cost function $\rcp$, under some admissible control
to be finite it is necessary and sufficient that
$(A,B)$ is stabilizable. 
If $(A,B)$ is stabilizable,
then $\Pi^{(n)}_{\alpha,t}$ converges as $n\to\infty$
to the unique positive definite solution
$\Pi^*_\alpha$ of the algebraic Riccati equation
\begin{equation}\label{E-Riccf}
\Pi_{\alpha}^* \,=\, R + \alpha A\transp \Pi_{\alpha}^* A 
- \alpha^2 A\transp \Pi_{\alpha}^* B 
(M+\alpha B\transp\Pi_{\alpha}^* B)^{-1}B\transp \Pi_{\alpha}^* A\,.
\end{equation} 
Consider the running cost
\begin{equation*}
c^{\alpha}_{q}(s,\Sigma) \,\df\,
\rcnt(s,q) + \trace\bigl(\widetilde\Pi_{\alpha}\Sigma\bigr)\,,\quad (s,\Sigma)\in\cZ\,,
\ t=0,\dotsc,n\,,
\end{equation*}
with
\begin{equation*}
\widetilde\Pi_{\alpha} \,\df\, R - \Pi^*_{\alpha}
 + \alpha A\transp \Pi^*_{\alpha} A\,.
\end{equation*}

We have the following theorem.

\medskip
\begin{theorem}\label{EnT3.1}
Suppose that $(A,B)$ is stabilizable.
Then, provided that $J_{\alpha}^*$, $\alpha\in(0,1)$, is locally bounded,
it takes the form
$J_{\alpha}^* (x,s,\Sigma)=
 \widetilde{J}_{\alpha}(x,\Sigma)
+ f^*_\alpha(s,\Sigma)$,
where
\begin{equation}\label{E-tJa}
\widetilde{J}_{\alpha}(x,\Sigma)\,\df\,
x\transp\Pi^*_{\alpha} x + \trace(\Pi^*_{\alpha} \Sigma)+
\frac{\alpha}{1-\alpha}\trace(\Pi^*_{\alpha} DD\transp)\,,
\end{equation}
and $f^*_\alpha(s,\Sigma)$ is the unique nonnegative
lsc solution of
the dynamic programming equation
\begin{equation}\label{EnT3.1A}
f^*_{\alpha}(s,\Sigma)\,=\,\min_{q\in\QSp}\,
\bigl\{c^{\alpha}_{q}(s,\Sigma)
+ \alpha \widehat\cT_{q} f^*_{\alpha}(s,\Sigma)\bigr\}\,,
\end{equation}
satisfying
\begin{equation}\label{EnT3.1B}
f^*_{\alpha}(s,\Sigma)
\,\le\, \kappa\bigl(1+\trace(\widetilde\Pi_{\alpha}\Sigma)\bigr)
\end{equation}
for some constant $\kappa=\kappa(\alpha)$.
Moreover, $f^*_\alpha$ is concave.
In addition, the stationary Markov control $(u^*_\alpha,q^*_\alpha)$, where
\begin{equation*}
\begin{split}
 u^*_\alpha(\widehat{X}_t) &\,=\, - K_{\alpha}^*\widehat{X}_{t}\,, \\
K_{\alpha}^*&\,=\, (M+\alpha B\transp \Pi_{\alpha}^* B)^{-1}
\alpha B\transp \Pi_{\alpha}^* A\,,
\end{split}
\end{equation*}
with $\Pi_{\alpha}^*$ as in \cref{E-Riccf},
and $q^*_\alpha$ is a selector of the minimizer of \cref{EnT3.1A}
is optimal for the infinite horizon discounted criterion.
\end{theorem}

\begin{IEEEproof}
Let $\Pi^{(n)}_{\alpha,t}(\Pi_\mathsf{fin})$ denote the solution of
\cref{E-Kf} satisfying $\Pi^{(n)}_{\alpha,n}= \Pi_\mathsf{fin}$,
and define $\widetilde\Pi^{(n)}_{\alpha,t}(\Pi_\mathsf{fin})$
analogously to \cref{E-tPiat}.
Recall the definitions in \cref{E-tJan,ET3.1B}.
Since $J^*_{\alpha,n}$ in \cref{ET3.1B} is optimal for the $n$-horizon problem,
we have
\begin{equation}\label{PnT3.1A}
\begin{aligned}
 \widetilde{J}_{\alpha,n} (x,\Sigma,0)
&+ \inf_{Q\in\widehat\Uadm}\,\widehat\Exp^{Q}_{s,\Sigma}\Biggl[\sum_{t=0}^{m-1}\alpha^{t}
\Bigl(\rcnt(S_t,Q_t)
+\trace\bigl(\widetilde\Pi^{(n)}_{\alpha,t}(0) \widehat\Pi_t\bigr)\Bigr)\Biggr]\\
&\,\le\, J^*_{\alpha,n}(x,s,\Sigma,0)
\,\le\, J_{\alpha}^*(x,s,\Sigma)\qquad\text{for all\ } m\le n\,.
\end{aligned}
\end{equation}
Now, since $\widetilde\Pi^{(n)}_{\alpha,t}(0)\to \widetilde\Pi_\alpha$
as $n\to\infty$ for each $t\in\NN_0$, we have
\begin{equation}\label{PnT3.1B}
\widehat\Exp^{Q}_{s,\Sigma}\Biggl[\sum_{t=0}^{m-1}\alpha^{t}
\trace\bigl(\widetilde\Pi^{(n)}_{\alpha,t}(0) \widehat\Pi_t\bigr)\Biggr]
\,\xrightarrow[]{n\to\infty}\,
\widehat\Exp^{Q}_{s,\Sigma}\Biggl[\sum_{t=0}^{m-1}\alpha^{t}
\trace\bigl(\widetilde\Pi_\alpha \widehat\Pi_t\bigr)\Biggr]
\end{equation}
for each fixed $m\le n$ and $Q\in\widehat\Uadm$, and also that
$\lim_{n\to\infty}\widetilde{J}_{\alpha,n}(x,\Sigma,0)=\widetilde{J}_{\alpha}(x,\Sigma)$.
Thus, by \cref{E-tJan,PnT3.1A,PnT3.1B}, we have
\begin{equation}\label{PnT3.1C}
\widetilde{J}_{\alpha}(x,\Sigma) +
\inf_{Q\in\widehat\Uadm}\,\widehat\Exp^{Q}_{s,\Sigma}\Biggl[\sum_{t=0}^{m-1}\alpha^{t}
c^{\alpha}_{Q_t}(S_t,\widehat\Pi_t)\Biggr]
\,\le\, \lim_{n\to\infty}\,J^*_{\alpha,n}(x,s,\Sigma,0)\,\le\, J_{\alpha}^*(x,s,\Sigma)
\end{equation}
for all $m\in\NN$.
Let
\begin{equation}\label{PnT3.1D}
\Bar{f}_n(s,\Sigma) \,\df\, \inf_{Q\in\widehat\Uadm}\,\widehat\Exp^{Q}_{s,\Sigma}
\Biggl[\sum_{t=0}^{m-1}\alpha^{t}
c^{\alpha}_{Q_t}(S_t,\widehat\Pi_t)\Biggr]\,.
\end{equation}
It follows from the theory of dynamic programming that the sequence
$\{\Bar{f}_n\}_{n\in\NN}$ is the solution of the equation
\begin{equation*}
\Bar{f}_{n+1}(s,\Sigma) \,=\, \min_{q\in\QSp}\,\bigl\{c^{\alpha}_{q}(s,\Sigma)
 + \alpha \widehat\cT_{q} \Bar{f}_{n}(s,\Sigma) \bigr\}\,,\quad (s,\Sigma)\in\cZ\,,
\end{equation*}
with $\Bar{f}_0\equiv0$.
It is clear that, for each $n\in\NN$,
$\Bar{f}_{n}(s,\Sigma)=g_n(s,\widetilde\Pi_\alpha\Sigma)$,
for some nonnegative function $g_n$, which
by \cref{L2.1}, is concave, lower semicontinuous, and nondecreasing
in its second argument (with respect to $\preceq$).
Since
$\widetilde{J}_{\alpha}+\Bar{f}_n\le{J}^*_{\alpha}$ by \cref{PnT3.1C},
and $J^*_\alpha$ is locally bounded,
$\Bar{f}_n$ converges as $n\to\infty$ to a some nonnegative
lsc function $f^*_\alpha$ which takes the form $g(s,\widetilde\Pi_\alpha\Sigma)$,
with $g$ concave in its second argument.
Therefore, the bound \cref{EnT3.1B} holds for some positive constant
$\kappa$ depending on $\alpha$.

Now, consider $J^*_{\alpha,n}(x,s,\Sigma,\Pi^*_\alpha)$,
and note that by setting $\Pi_{\mathsf{fin}}=\Pi_{\alpha}^*$, the backward recursion in
\cref{E-Kf} is $t$-invariant.
It is straightforward to show,
using \cref{E-tJan} and by the optimality of $J^*_\alpha$, that
\begin{equation}\label{PnT3.1E}
\widetilde{J}_{\alpha}(x,\Sigma) +
f^*_\alpha(s,\Sigma)\,=\,\lim_{n\to\infty} J^*_{\alpha,n}(x,s,\Sigma,\Pi^*_\alpha)
\,\ge\,
J^*_\alpha(x,s,\Sigma)\,,
\end{equation}
and since the reverse inequality was shown, the characterization of $J^*_\alpha$ in
\cref{E-tJa,EnT3.1A} is established.

Let $q^*_\alpha$ be a selector from the minimizer of \cref{EnT3.1A}.
Then we have
\begin{equation}\label{PnT3.1F}
f^*_\alpha(s,\Sigma) \,=\,
\widehat\Exp^{q^*_\alpha}_{s,\Sigma}\Biggl[\sum_{t=0}^{n-1}\alpha^{t}
c^{\alpha}_{q^*_\alpha}(S_t,\widehat\Pi_t)\Biggr]
+ \alpha^n \widehat\Exp^{q^*_\alpha}_{s,\Sigma}
\bigl[f^*_\alpha(S_n,\widehat\Pi_n)\bigr]\,.
\end{equation}
Since $f^*_\alpha$ is locally bounded, 
the first term on the right hand side is finite, and so
$\alpha^{n}\widehat\Exp^{q^*_\alpha}_{s,\Sigma} \bigl[
c^{\alpha}_{q^*_\alpha}(S_n,\widehat\Pi_n)\bigr] \to 0$ as $n \to \infty$.
Hence, since $c^{\alpha}_q(s,\Sigma)\ge\trace(\widetilde\Pi_{\alpha}\Sigma)$,
the second term on the right hand side must vanish as $n\to\infty$
by \cref{EnT3.1B}.
Thus, taking limits in \cref{PnT3.1F}, using monotone convergence,
we have
\begin{equation}\label{PnT3.1G}
f^*_\alpha(s,\Sigma) \,=\,
\widehat\Exp^{q^*_\alpha}_{s,\Sigma}\Biggl[\sum_{t=0}^{\infty}\alpha^{t}
c^{\alpha}_{q^*_\alpha}(S_t,\widehat\Pi_t)\Biggr]
\,\le\, \inf_{Q\in\widehat\Uadm}\,\widehat\Exp^{Q}_{s,\Sigma}
\Biggl[\sum_{t=0}^{\infty}\alpha^{t}
c^{\alpha}_{Q_t}(S_t,\widehat\Pi_t)\Biggr]\,,
\end{equation}
where for the inequality we use \cref{PnT3.1D}.
This together with \cref{E-sepA,PnT3.1E,PnT3.1G} implies that
\begin{equation*}
\lim_{n\to\infty} J^*_{\alpha,n}(x,s,\Sigma,\Pi^*_\alpha) \,=\,
\widehat\Exp^{q^*_\alpha,u^*_\alpha}_{s,\Sigma}\Biggl[\sum_{t=0}^{\infty}\alpha^{t}
\Bigl(\rcnt(S_t,Q_t) + \rcp(X_{t},U_{t})\Bigr)\Biggr]\,.
\end{equation*}
This establishes the optimality of $(q^*_\alpha,u^*_\alpha)$.

It remains to show uniqueness of the solutions
to \cref{EnT3.1A} in the class of functions stated
in the theorem.
Suppose that $V\colon\psdef\to\RR$
is a nonnegative lsc function
 which satisfies \cref{EnT3.1A,EnT3.1B} for some constant $\kappa$.
Then,
$V(s,\Sigma)\le c^{\alpha}_{q^*_\alpha}(s,\Sigma)
+ \alpha \widehat\cT_{q^*_\alpha} V(s,\Sigma)$,
from which we deduce that
\cref{PnT3.1F} holds with inequality if we replace $f^*_\alpha$ with $V$.
Since $f^*_\alpha(s,\Sigma)\ge\trace(\widetilde\Pi_\alpha\Sigma)$ by \cref{EnT3.1A},
we must have $\alpha^n \widehat\Exp^{q^*_\alpha}_{s,\Sigma}
\bigl[V(S_n,\widehat\Pi_t)\bigr] \to 0$ as $n \to \infty$ by the argument used earlier.
Therefore, $V\le f^*_\alpha$ on $\cZ$.
On the other hand, if $\Hat{q}$ is a selector from the minimizer
of the dynamic programming equation
of $V(s,\Sigma)=\min_{q\in\QSp}\,
\bigl\{c^{\alpha}_{q}(s,\Sigma)
+ \alpha \widehat\cT_{q} V(s,\Sigma)\bigr\}$,
it readily follows as in \cref{PnT3.1F} that
\begin{equation*}
V(s,\Sigma) \,\ge\, \widehat\Exp^{\Hat{q}}_{s,\Sigma}\Biggl[\sum_{t=0}^{\infty}\alpha^{t}
c^{\alpha}_{\Hat{q}}(S_t,\widehat\Pi_t)\Biggr]
\,\ge\, f^*_\alpha(s,\Sigma)
\end{equation*}
since $f^*_\alpha$ is the optimal value function by \cref{PnT3.1G}.
Therefore, $V=f^*_\alpha$, and this completes the proof.
\end{IEEEproof}

\subsection{The ergodic control problem}

As mentioned earlier $(A,B)$ stabilizable is a necessary condition for
the ergodic cost problem to be well posed.
So this is assumed throughout the rest of the paper without further mention,
except for emphasis.
Next, consider the estimation part of the problem.
As we have seen in \cref{S2A}, this corresponds to
an optimal control problem for
the controlled Markov chain $\{Z_t\}_{t\in\NN_0}$ with
transition kernel given in \cref{E-cT}.
Recall that $\widehat\Exp^Q_z$ denotes the expectation operator
on the path space of this chain, in analogy to the definition in \cref{E-hExp}.
Note that the running cost for the finite horizon problem takes
the form in \cref{E-caqt}, and that $\widetilde\Pi^{(n)}_{\alpha,t}$ converges
to $\widetilde\Pi_\alpha$ as $n\to\infty$ for any fixed $t$.
In turn $\widetilde\Pi_\alpha$ converges to the
solution $\widetilde\Pi^*$ of the algebraic Riccati equation
in \cref{ED3.2A} as $\alpha\to1$.
Consequently, for the ergodic control problem to be well posed,
there must exist an admissible policy $\{Q_t\}_{t\in\NN_0}$
such that
\begin{equation}\label{E-stab}
\limsup_{T\to\infty}\,\frac{1}{T}\,
\sum_{t=0}^{T-1}\widehat\Exp^{\,Q}_z\bigl[\trace(\widetilde\Pi^*\widehat\Pi_t)\bigr]
\,<\,\infty \quad\forall z\in\cZ\,.
\end{equation}
This also renders the discounted value function locally bounded, since
by applying a well known Tauberian theorem \cite[Theorem~2.2]{Sz-Fil},
and using the convergence of $\widetilde\Pi_\alpha$ to $\widetilde\Pi^*$,
we obtain
\begin{equation}\label{E-tauber}
\limsup_{\alpha\nearrow1}\, (1-\alpha)\,f^*_\alpha (z) \,\le\,
\limsup_{T\to\infty}\,\frac{1}{T}\,
\sum_{t=0}^{T-1}\widehat\Exp^{\,Q}_z\bigl[c_Q(S_t,\widehat\Pi_t\bigr]
\,<\,\infty\,,
\end{equation}
with $c_q$ defined as in \cref{ED3.2C}.
Therefore, \cref{E-stab} is a necessary condition and cannot be avoided.
In the absence of intermittency it has been shown in \cite{Wu2008}
that a necessary and sufficient condition for \cref{E-stab} to hold
is that $(\overline{C},A)$ is detectable, where
$\overline{C}\df [C_{q_{1}}\transp\,|\,\cdots\,|\,C_{q_{\abs{\QSp}}}\transp]\transp$.
However, with intermittency, the process $\{\widehat\Pi_t\}_{t\in\NN_0}$
is modulated by the Markov chain $\{S_t\}_{t\in\NN_0}$, and therefore switches
among many environments.
Simple algebraic conditions for stability of this process do not seem possible,
even in the uncontrolled setting without sensor scheduling \cite{Sinopoli2004}.
Therefore, in this work, we simply assume that the estimation is 
stabilizable under some scheduling policy, otherwise the ergodic control problem
is not well posed.
We summarize our hypotheses below.

\medskip
\begin{assumption}\label{A3.1}
The following hold:
\begin{itemize}
\item[(i)]
There exists $(s_\circ,\Sigma_\circ)\in\cZ$,
and an admissible query process $\widetilde{Q}=\{\widetilde{Q}_{t}\,\colon t\ge 0\}$ 
such that
\begin{equation}\label{EA3.1A}
\limsup_{T\to\infty}\,\frac{1}{T}\,
\sum_{t=0}^{T-1}\widehat\Exp^{\,\widetilde{Q}}_{s_\circ,\Sigma_\circ}\bigl[\trace(\widetilde\Pi^*\widehat\Pi_t)\bigr]
\,<\,\infty\,.
\end{equation}

\item[(ii)]
The pair $(A,D)$ is controllable, and $(A,B)$ is stabilizable.

\item[(iii)]
The controlled {M}arkov chain governing the network,
with transition matrix $P_q= [p_q(s,s')]_{s,s'\in\SSp}$ given in \cref{E-net},
has the following property:
There exists $\vartheta>0$ and an irreducible stochastic matrix
$\widetilde{P}=[\Tilde{p}(s,s')]_{s,s'\in\SSp}$ such that
$p_q(s,s')\ge \vartheta\Tilde{p}(s,s')$ for all $s,s'\in\SSp$
and all $q\in\QSp$.
\end{itemize}
\end{assumption}

We remark on parts (ii) and (iii) of \cref{A3.1}.
As we show in \cref{L3.1}, under (ii) the support
of the $(d-1)^{\mathrm{th}}$ iterate of $\widehat\cT_q$
lies in a subset of $\pdef$ whose eigenvalues are bounded away
from $0$ under any policy.
Part (iii) is an `irreducibility' assumption.
Note that \cref{A3.1}\,(i) assumes that the average cost
is finite for some initial condition.
As we show later, this together with parts (ii) and (iii), implies \cref{E-stab}.

\subsection{Structural properties of the value function}

Note that in the proof of \cref{EnT3.1} we showed that
$f_{\alpha}^*(s,\cdot)$ is non-decreasing, so
\begin{equation*}
\inf_{\Sigma\in\psdef}f_{\alpha}^*(s,\Sigma) \,=\, f_{\alpha}^*(s,0)\,.
\end{equation*}

\begin{definition}\label{D3.1}
For a set $G\in\psdef$ we define
\begin{align*}
\spn_{G}\,f_{\alpha}^*(s,\cdot) & \,\df\, 
\sup_{\Sigma\in G}\, f_{\alpha}^*(s,\Sigma)
 - \inf_{\Sigma\in G}\, f_{\alpha}^*(s,\Sigma)\,, \\
\spn_{\SSp\times G}\,f_{\alpha}^* & \,\df\, 
\sup_{(s,\Sigma)\in\SSp\times G}\, f_{\alpha}^*(s,\Sigma)
 - \inf_{(s,\Sigma)\in\SSp\times G}\, f_{\alpha}^*(s,\Sigma)\,.
\end{align*}
We also define the sets
\begin{equation}\label{E-sets}
\begin{aligned}
\pdef_\varepsilon &\,\df\, \{\widehat\Pi\in\pdef\,\colon
\underline{\sigma}(\widehat\Pi)>\varepsilon\}\,,\\
\cB_{r} &\,\df\, \{\Sigma\in\psdef \colon \trace(\Sigma)\le r\}\,,
\quad r>0\,.
\end{aligned}
\end{equation}
We let $\uptau(G)$ denote the first exit time from a set $G\subset\cZ$,
and adopt the following shortened notation:
\begin{equation}\label{E-hit}
\uptau_{r} \,\df\, \uptau(\SSp\times\cB^c_{r})\,,
\quad\text{and\ } \Breve\uptau_r \,\df\, \uptau(\SSp\times\cB^c_r)\,,
\end{equation}
\end{definition}

\begin{lemma}\label{L3.1}
Provided the pair $(A,D)$ is controllable, 
there exists a constant  $\varepsilon_\circ >0$
such that for any
admissible querying policy $Q$ we have
\begin{equation*}
\Prob^{Q}_{s,\Sigma}(\widehat\Pi_t\in\pdef_{\varepsilon_\circ})\,=\,1
\quad\forall\,t\ge d-1\,,\ 
\forall\,(s,\Sigma)\in\SSp\times\psdef\,.
\end{equation*}
\end{lemma}

\begin{IEEEproof}
This is an adaptation of the proof of \cite[Lemma 3.5]{Wu2008}.
Note that for any sample path string
$\widehat\Pi_t, \widehat\Pi_{t+1},\dotsc,\widehat\Pi_{t+d}$,
with $t\ge0$, we have either $\widehat\Pi_{k+1} = \Xi(\widehat\Pi_{k})$,
or $\widehat\Pi_{k+1} = \cT_q(\widehat\Pi_{k})$ for some $q\in\QSp$.
If we write $\cT_q$ as
\begin{equation}\label{PL3.1A}
\cT_{q}(\widehat\Pi) \,=\,
\bigl(I -\widehat{K}_{q,1}(\widehat\Pi)C_{q}\bigr)\Xi(\widehat\Pi)
\bigl(I -\widehat{K}_{q,1}(\widehat\Pi)C_{q}\bigr)\transp
+\widehat{K}_{q,1}(\widehat\Pi) F_q F_q\transp \widehat{K}_{q,1}\transp(\widehat\Pi)\,.
\end{equation}
Then, as argued in the proof of \cite[Lemma 3.5]{Wu2008} using \cref{PL3.1A},
we have
\begin{equation}\label{PL3.1B}
\begin{aligned}
\ker\bigl(\cT_{q}(\widehat\Pi)\bigr) &\,=\, \ker\bigl(\Xi(\widehat\Pi)\bigr)\\
&\,=\, \ker(A\widehat\Pi A\transp)\cap\ker (DD\transp)\\
&\,=\, \ker(\widehat\Pi A\transp)\cap\ker (D\transp)\,.
\end{aligned}
\end{equation}
Iterating \cref{PL3.1B}, and
with $\cT_{\bm q}^{d}$ denoting the composition
$\cT_{q_d}\circ\dotsb\circ\cT_{q_1}$, we have
\begin{equation*}
\ker\bigl(\cT_{\bm q}^{d}(\widehat\Pi)\bigr) \,\subseteq\,
\ker(D\transp)\cap\cdots\cap
\ker\bigl(D\transp (A\transp)^{d-1}\bigr)\,.
\end{equation*}
Since $(A,D)$ is controllable, we obtain
$\ker(\widehat\Pi_{t+d})=\ker\bigl(\cT_{\bm q}^{d}(\widehat\Pi_t)\bigr)=\{0\}$.
Using the monotonicity property in \cref{L2.1}, it is clear
that the smallest eigenvalue of $\cT_{\bm q}^{d}(\widehat\Pi_t)$ is bounded below
by that of the matrix $\cT_{\bm q}^{d}(0)$, that is,
 obtained over the same sample path string
starting with $\widehat\Pi_t=0$.
Choosing $\varepsilon_\circ$ 
to be the minimal eigenvalue of $\cT_{\bm q}^{d}(0)$, over
the finitely many $d$-step  sequences of queries
and network states, the result follows.
\end{IEEEproof}

Combining \cref{EA3.1A}, \cref{L3.1}, \cref{A3.1}\,(iii),
and the monotonicity of the map $\widehat\cT_{q}$ in \cref{L2.1},
we have the following corollary.
\medskip
\begin{corollary}\label{C3.1}
Under \cref{A3.1}, there exists an admissible query process $Q=\{Q_{t}\,\colon t\ge 0\}$
such that
\begin{equation*}
\limsup_{T\to\infty}\,\frac{1}{T}\,
\sum_{t=0}^{T-1}\widehat\Exp^{\,Q}_{s,\Sigma}\bigl[\trace(\widetilde\Pi^*\widehat\Pi_t)\bigr]
\,<\,\infty\qquad \forall (s, \Sigma) \in\cZ\,.
\end{equation*}
\end{corollary}
\begin{IEEEproof}
Using \cref{L2.1} and noting that Trace is convex and non-decreasing, for any constant $m>1$, a query $q\in\QSp$, and a $z = (s,\Sigma)\in\cZ$, we have
\begin{equation*}
\widehat\cT_{q} \trace(\widetilde\Pi^*\Sigma) \,\le\, \Bigl(1 - \frac{1}{m}\Bigr) \widehat\cT_{q} \trace(0) + \frac{1}{m} \widehat\cT_{q} \trace(m \widetilde\Pi^* \Sigma).
\end{equation*} 
Rearranging and iterating for a sequence of sensor queries $\{q_0,\dots, q_k\}$ yields
\begin{equation}\label{E-bd1}
\widehat\cT_{q_k} \circ \dots \circ \widehat\cT_{q_0} \trace(m\widetilde\Pi^* \Sigma) \,\ge\, m \widehat\cT_{q_k}\circ \dots \circ \widehat\cT_{q_0} \trace(\widetilde\Pi^*\Sigma).  
\end{equation}

By the monotonicity of trace and the
previously shown properties of $\widehat{\cT}$, 
\cref{EA3.1A} also holds for an initial condition $(s_\circ,0)$ and a query process $\widetilde{Q}$ where $s_\circ$ and $\widetilde{Q}$ are as in \cref{A3.1} (i).
Using \cref{A3.1}\,(ii) and \cref{L3.1}, the eigenvalues of $\widehat{\Pi}_{t}$ 
are bounded away from $0$ by some $\varepsilon_\circ>0$ for all $t\ge d$   when starting with an initial covariance $0$. 
Combining this result with \cref{A3.1}\,(iii),
it follows that for any $s\in\SSp$
there 
is an $n\in\NN$ and 
a strictly positive definite $\Sigma$ such that
\Cref{EA3.1A} holds for $(s,\Sigma)$ with a 
query sequence $\widetilde{Q}'$ given
by $\widetilde{Q}'_{t}=\widetilde{Q}_{t+n}$.
Let $\varepsilon>0$ denote the smallest of the eigenvalues of the
$\Sigma$ matrices. 
Again using monotonicity of trace and the properties of $\widehat{\cT}$,
if $\Sigma\in\psdef$ such that $\trace(\Sigma)<\varepsilon$ then for any $s\in\SSp$
\Cref{EA3.1A} holds for $(s,\Sigma)$ and $\widetilde{Q}'$.

Then for any $(s,\Sigma)\in\cZ$, 
choose $\varepsilon'$ such that
$0<\varepsilon'<\dfrac{\varepsilon}{\trace(\Sigma)}$.
For any $t>0$, using \cref{E-bd1},
\begin{equation}\label{E-aux1}
\varepsilon' \,\Exp_{s,\Sigma}^{\,\widetilde{Q}'}
\bigl[\trace(\widetilde \Pi^*\widehat\Pi_t)\bigr]
\,\le\,
\Exp_{s,\varepsilon' \,\Sigma}^{\,\widetilde{Q}'}
\bigl[\trace(\widetilde \Pi^*\widehat\Pi_t)\bigr]\,.
\end{equation}
Note that \Cref{EA3.1A} holds for the right had side of \cref{E-aux1}
for any $(s,\Sigma)\in\cZ$ and a query process $\widetilde{Q}'$
since $\trace(\varepsilon' \Sigma) < \varepsilon$ which concludes the proof.
\end{IEEEproof}

\begin{lemma}\label{L3.2}
The following hold:
\begin{itemize}
\item[(a)]
There exists a positive constant $\kappa_0$, such that the function
\begin{equation*}
\Bar{f}_{\alpha,s}(\Sigma) \,\df\, f_{\alpha}^*(s,\Sigma) - f_{\alpha}^*(s,0)\,,
\quad s\in\SSp\,,
\end{equation*}
satisfies the bound
\begin{equation}\label{EL3.2A}
\Bar{f}_{\alpha,s}(\Sigma) \,\le\,
\kappa_0\bigl(1+\trace(\widetilde\Pi_\alpha\Sigma)\bigr)
\quad\forall\, (s,\Sigma)\in\cZ\,,
\end{equation}
and for all $\alpha\in (0,1)$.
\item[(b)]
There exists a constant $\varrho^*$ such that
$\limsup_{\alpha\nearrow1}\, (1-\alpha) f^*_\alpha(s,0) =\varrho^*$
for all $s\in\SSp$.
\item[(c)]
In addition, the family of functions
$\{\Bar{f}_{\alpha,s}\,\colon\alpha\in (0,1),\,s\in\SSp\}$
is locally Lipschitz equicontinuous.
\end{itemize}
\end{lemma}

\begin{IEEEproof}
Recall \cref{D3.1}.
Let $\Tilde{r}>0$ be a constant such that all sample paths of length
$2d$ of the process $\widehat\Pi_t$ in \cref{E-fltr3} starting
from $\widehat\Pi_0=0$ are contained in $\cB_{\Tilde{r}}$, and
fix $\pdef_{\varepsilon_\circ}$ as in \cref{L3.1}.
Let $\varepsilon=\frac{\varepsilon_\circ}{\Tilde{r}}$.
The proof relies on the following simple convexity argument:
if $\Sigma'\in\pdef_{\varepsilon_\circ}\cap \cB_{\Tilde{r}}$ then
\begin{equation*}
\Sigma'' \,\df\, \frac{\Sigma'-\varepsilon\Sigma}{1-\varepsilon}\,\in\,
\cB_{\Tilde{r}}\quad \forall\,\Sigma\in\cB_{\Tilde{r}}\,.
\end{equation*}
Since $f^*_\alpha$ is concave, this implies that
\begin{equation}\label{PL3.2A}
\begin{aligned}
f^*_\alpha(s,\Sigma) - f^*_\alpha(s,\Sigma')
&\,\le\, (1-\varepsilon)\bigl(f^*_\alpha(s,\Sigma) - f^*_\alpha(s,\Sigma'')\bigr)\\
&\,\le\, (1-\varepsilon)\,\spn_{\cB_{\Tilde{r}}}\,f_{\alpha}^*(s,\cdot)
\quad \forall\,\Sigma\in\cB_{\Tilde{r}}\,,\ \forall\,s\in\SSp\,.
\end{aligned}
\end{equation}
By \cref{PnT3.1F}, we have
\begin{equation}\label{PL3.2B}
f^*_{\alpha}(s,0) \,=\, H_\alpha(s,0) +
\frac{1}{d} \sum_{t=d}^{2d-1} \alpha^{t}\Exp_{s,0}^{q_{\alpha}^*}
\bigl[f^*_{\alpha}(S_{t},\widehat\Pi_t)\bigr]\,,
\end{equation}
where $H_\alpha(s,0)$ is a bounded function.
With $P_{q^*_\alpha}^{(k)}$ denoting the the $k^{\mathrm th}$ composition
of the network transition matrices under the policy $q^*_\alpha$,
we define the substochastic matrix
$\Bar{P}(\alpha)=\bigl[\Bar{p}_{ij}(\alpha)\bigr]_{i,j\in\SSp}$ by
\begin{equation*}
\Bar{P}(\alpha)\,\df\, \frac{1}{d} \sum_{t=d}^{2d-1} \alpha^{t}\Exp_{s,0}^{q_{\alpha}^*}
\bigl[P_{q^*_\alpha}^{(t)}\bigr]\,.
\end{equation*}
It is clear from \cref{E-cT} that
\begin{equation}\label{PL3.2C}
\frac{1}{d} \sum_{t=d}^{2d-1} \alpha^{t}\Exp_{s,0}^{q_{\alpha}^*}
\bigl[\indic{\{S_t=s'\}}\bigr]\,=\, \Bar{p}_{s,s'}(\alpha)\,.
\end{equation}
By \cref{A3.1}\,(ii), $\Bar{P}(\alpha)$ is a positive matrix,
and all its elements are bounded below away from $0$ for all $\alpha\ge\nicefrac{1}{2}$,
and thus, as is well known, it has a unique positive eigenvector $\uppi(\alpha)$
corresponding to its  eigenvalue of largest modulus $\rho(\alpha)$.
Let $\xi_s^\alpha\in\cB_{\Tilde{r}}$ be a point where
$f^*_\alpha(s,\cdot)$ attains its maximum.
It follows from \cref{PL3.2A,PL3.2C} that
\begin{equation}\label{PL3.2D}
\Bar{p}_{s,s'}(\alpha) f_{\alpha}^*(s',\xi_{s'}^\alpha)-
\frac{1}{d} \sum_{t=d}^{2d-1} \alpha^{t}\Exp_{s,0}^{q_{\alpha}^*}
\bigl[f^*_{\alpha}(S_{t},\widehat\Pi_t)\indic{\{S_t=s'\}}\bigr]
\,\le\, (1-\varepsilon)\,\Bar{p}_{s,s'}(\alpha)\,
\spn_{\cB_{\Tilde{r}}}\,f_{\alpha}^*(s',\cdot)\,.
\end{equation}
Thus, suppressing the dependence  $\Bar{P}$, $\uppi$, $\rho$ and $\xi$ on $\alpha$,
in order to simplify the notation, and denoting the network states
as $i,j\in\SSp$, we obtain from \cref{PL3.2B,PL3.2D} that
\begin{equation}\label{PL3.2E}
\begin{aligned}
\sum_{i\in\SSp} \uppi_i \Biggl(\sum_{j\in\SSp} \Bar{p}_{ij}
f_{\alpha}^*(j,\xi_j)-f_{\alpha}^*(i,0)\Biggr)
&\,\le\, \sum_{i\in\SSp} \uppi_i \sum_{j\in\SSp} 
(1-\varepsilon)\,\Bar{p}_{ij}\spn_{\cB_{\Tilde{r}}}\,f_{\alpha}^*(j,\cdot)\\
&\,=\, \rho (1-\varepsilon)\sum_{j\in\SSp} \uppi_j
\spn_{\cB_{\Tilde{r}}}\,f_{\alpha}^*(j,\cdot)\,.
\end{aligned}
\end{equation}
The left-hand side of \cref{PL3.2E} evaluates to
\begin{equation}\label{PL3.2F}
\rho \sum_{j\in\SSp} \uppi_j\,f_{\alpha}^*(j,\xi_j)-
\sum_{j\in\SSp} \uppi_j\,f_{\alpha}^*(j,0)
\,=\, \rho \sum_{j\in\SSp} \uppi_j\,\spn_{\cB_{\Tilde{r}}}\,f_{\alpha}^*(j,\cdot)
-(1-\rho)\sum_{j\in\SSp} \uppi_j\,f_{\alpha}^*(j,0)\,.
\end{equation}
Therefore, by \cref{PL3.2E,PL3.2F}, we have
\begin{equation}\label{PL3.2G}
\sum_{j\in\SSp} \uppi_j\,\spn_{\cB_{\Tilde{r}}}\,f_{\alpha}^*(j,\cdot)
\,\le\,\frac{1-\rho}{\varepsilon}
 \sum_{j\in\SSp} \uppi_j\,f_{\alpha}^*(j,0)\,.
\end{equation}
Considering the right eigenvector $(1,\dotsc,1)$ of $\Bar{P}$,
an easy calculation shows that
$\rho = \frac{1}{d}\alpha^d (1+\dotsb+\alpha^{d-1})\ge \alpha^{2d-1}$.
Therefore,
\begin{equation}\label{PL3.2H}
1-\rho \,\le\, 2d(1-\alpha)
\end{equation}
\Cref{PL3.2G,PL3.2H} together with an application of the
Tauberian theorem as in \cref{E-tauber}, using \cref{C3.1}, show that
\begin{equation}\label{PL3.2I}
\begin{aligned}
\sum_{j\in\SSp} \uppi_j\,\spn_{\cB_{\Tilde{r}}}\,f_{\alpha}^*(j,\cdot)
\,<\,\infty\,.
\end{aligned}
\end{equation}
Then, \cref{EL3.2A} follows from \cref{PL3.2I} and concavity, and
this completes the proof of part (a).

We turn our attention now to part (b).
Using \cref{PL3.2B,PL3.2C}, adding and subtracting
terms, we write
\begin{equation*}
f^*_\alpha(i,0) \,\le\, H_\alpha(i,0)
+ \sum_{j\in\SSp} \Bar{p}_{ij}\,f_{\alpha}^*(j,0)
+ \sum_{j\in\SSp} \Bar{p}_{ij}\spn_{\cB_{\Tilde{r}}}\,f_{\alpha}^*(j,\cdot)\,,
\end{equation*}
Which we rearrange as
\begin{equation}\label{PL3.2J}
\sum_{j\in\SSp} \Bar{p}_{ij}\,\bigl(f^*_\alpha(i,0)-f_{\alpha}^*(j,0)\bigr)
\,\le\, H_\alpha(i,0)
+ \biggl(1-\sum_{j\in\SSp} \Bar{p}_{ij}\biggr)\,f_{\alpha}^*(i,0)
+ \sum_{j\in\SSp} \Bar{p}_{ij}\spn_{\cB_{\Tilde{r}}}\,f_{\alpha}^*(j,\cdot)\,,
\end{equation}
The first term on the right-hand side of \cref{PL3.2J} is bounded
as mentioned earlier, and the third term is bounded by \cref{PL3.2I}.
It is easy to see that
$1-\sum_{j\in\SSp} \Bar{p}_{ij}\sim 1-\alpha$.
A standard argument using
a well known Tauberian theorem \cite[Theorem~2.2]{Sz-Fil} together
with \cref{A3.1}\,(iii) then shows that
$\lim_{\alpha\nearrow1}\, (1-\alpha) f^*_\alpha(i,0)$ is finite for all
$i\in\SSp$, which implies that the 
the second term is also bounded.
Since $\{\Bar{p}_{ij}\}$ are strictly positive, and \cref{PL3.2J} holds
for all $i\in\SSp$, part (b) follows.

Concerning part (c), since the  functions $\{\Bar{f}_{\alpha,i}\,\colon
i\in\SSp, \alpha\in(\nicefrac{1}{2},1)\}$ are concave and locally
bounded by \cref{PL3.2I}, they are  Lipschitz equicontinuous
on any set of the form $\pdef_{\varepsilon}\cap \cB_{r}$,
with a Lipschitz constant which depends only on $\varepsilon>0$ and $r>0$
\cite[Theorem 10.6]{Rockafellar1970}.
Fix an initial $(s,\Sigma)\in\SSp\times\psdef$, and let 
$\bm{q}=\{q_0,\dotsc,q_{2d-1}\}$ be the sequence of the first
$2d-1$ queries from the 
$\alpha$-discounted optimal control $q^*_\alpha$.
We apply these in an open loop fashion to control
the process starting at $(s,\Sigma')$
From \cref{PnT3.1F}, and since $\bm{q}$ is suboptimal
for the initial condition $(s,\Sigma')$, we obtain
\begin{equation*}
\begin{aligned}
f^*_{\alpha}(s,\Sigma')-f^*_{\alpha}(s,\Sigma) &\,\le\,
H_\alpha(s,\Sigma') -H_\alpha(s,\Sigma)\\
&\mspace{100mu} +
\frac{1}{d} \sum_{t=d}^{d-1} \alpha^{t}\,
\bigl(\Exp_{s,\Sigma'}^{\bm{q}}
\bigl[f^*_{\alpha}(S_{t},\widehat\Pi_t)\bigr]
-\Exp_{s,\Sigma}^{q_{\alpha}^*}
\bigl[f^*_{\alpha}(S_{t},\widehat\Pi_t)\bigr]\bigr)\,,
\end{aligned}
\end{equation*}
It is clear that $\Sigma\mapsto H_\alpha(s,\Sigma)$ is locally Lipschitz.
Since the support of
$\Exp_{s,\Sigma}^{q_{\alpha}^*}
\bigl[f^*_{\alpha}(S_{t},\cdot)\bigr]$ lies in
$\pdef_{\varepsilon_\circ}$ by \cref{L3.1},
it also follows that there is a Lipschitz estimate for the second term.
Reversing the role of $\Sigma$ and $\Sigma'$, we get the opposite inequality,
thus proving part (c).
This completes the proof of the lemma.
\end{IEEEproof}

\subsection{The Bellman equation}

We use the vanishing discount approach to derive the optimality
equation for the average cost problem.
We fix a point $\uptheta=(s_\circ,0)\in\cZ$.
A critical result enabling the vanishing discount approach is the following
result.

\medskip
\begin{definition}\label{D3.2}
We let $\Pi^*\in\pdef$ denote unique solution of the algebraic Riccati equation
\begin{equation}\label{ED3.2A}
\Pi^* \,=\, R + A\transp \Pi^* A {-} A\transp
\Pi^* B (M+B\transp \Pi^* B)^{-1}B\transp \Pi^* A\,,
\end{equation}
and define
\begin{equation}\label{ED3.2B}
\begin{aligned}
\widetilde\Pi^* &\,\df\, R - \Pi^* + A\transp \Pi^* A\,, \\
K^* &\,\df\, (M+B\transp \Pi^* B)^{-1}B\transp \Pi^* A\,.
\end{aligned}
\end{equation}
We also let $u^*$ denote the stationary Markov control
$u^*(\widehat{X}_t) = - K^*\widehat{X}_t$, with $\widehat{X}_t$ as
in \cref{E-fltr1}.
Lastly, for $q\in\QSp$, we define
\begin{equation}\label{ED3.2C}
c_q(s,\Sigma) \,\df\,
\rcnt(s,q) + \trace\bigl(\widetilde\Pi^*\Sigma\bigr)\,,\quad (s,\Sigma)\in\cZ\,.
\end{equation}
\end{definition}

\medskip
\begin{theorem}\label{T3.2}
There exist a constant $\varrho^*$ and a continuous concave function
${f^*\colon\cZ\to\RR_+}$ that satisfy
\begin{equation}\label{ET3.2A}
f^*(z) + \varrho^*
=\min_{q\in\QSp}\,\bigl\{c_q(z) + \widehat\cT_{q} f^*(z) \bigr\}
\quad\forall\,z\in\cZ\,,
\end{equation}
with $c_q$ as in \cref{ED3.2C}.
Let $q^*\colon\SSp\times\psdef\to\QSp$ be a selector of the 
minimizer in \cref{ET3.2A}.
The following hold:
\begin{itemize}
\item[(a)]
There exists a positive constant $\kappa_0^*$ such that
\begin{equation}\label{ET3.2B}
f^*(s,\Sigma)\,\le\, \kappa_0^*\bigl(1+\trace(\widetilde\Pi^*\Sigma)\bigr)
\quad\forall\,(s,\Sigma)\in\cZ\,.
\end{equation}
Also $f^*$ satisfies the lower bound
\begin{equation}\label{ET3.2C}
f^*(s,\Sigma) \,\ge\, \trace(\widetilde\Pi^*\Sigma) - \varrho^*
\quad\forall\,(s,\Sigma)\in\cZ\,.
\end{equation}
It is also the case that $f^*$ is nondecreasing in its second argument.
\item[(b)]
There exists positive constants $\theta_1\in(0,1)$ and $\theta_2$ such that
\begin{equation}\label{ET3.2D}
\min_{q\in\QSp}c_q(z) \,\ge\, \theta_{1} f^*(z)-\theta_{2}
\quad\forall\,z\in\cZ\,.
\end{equation}
In particular, we have
\begin{equation}\label{ET3.2E}
\widehat\cT_{q^*} f^*(z) \,\le\, \rho f^*(z) +(\rho^*+\theta_2)
\quad\forall\,z\in\cZ\,,
\end{equation}
with $\rho\df (1-\theta_1) <1$, which implies that
\begin{equation}\label{ET3.2F}
\Exp^{q^*}_z \bigl[f^*(Z_n)\bigr] \,\le\, \frac{\rho^*+\theta_2}{\theta_1}
+ \rho^n f^*(z)\quad\forall\,z\in\cZ\,,\ \forall\,n\in\NN\,.
\end{equation}
\item[(c)]
The stationary Markov control
$v^*=(q^*,u^*)$, with $u^*$ as in \cref{D3.2},
 is optimal for the average cost criterion, and 
\begin{equation*}
J^*\,=\,\varrho^* + \trace(\Pi^* DD\transp)\,.
\end{equation*}
\end{itemize}
\end{theorem}

\begin{IEEEproof}
Since the system is stabilizable, the Riccati equation \cref{E-Riccf}
converges as $\alpha\to 1$ to \cref{ED3.2A}
which has a unique solution $\Pi^*\in\pdef$.
The feedback given by \cref{ED3.2B} is then optimal for any given
querying sequence, and we only need consider optimal sensor scheduling.

Fix a point $\uptheta\df(s_\circ,0)\in\cZ$ and let
$\Bar{f}_\alpha(z) \df f^*_\alpha(z) - f^*_\alpha(\uptheta)$.
Then write \cref{EnT3.1A} as
\begin{equation}\label{PT3.2A}
\Bar{f}_{\alpha}(s,\widehat\Pi)\,=\,\min_{q\in\QSp}\,
\bigl\{c^\alpha_q(z) + \alpha \widehat\cT_{q} f^*_{\alpha}(s,\widehat\Pi)\bigr\}
+ (1-\alpha)f^*_\alpha(\uptheta) \,.
\end{equation}
Applying \cref{L3.2}, we can take limits in \cref{PT3.2A}
along some subsequence $\alpha_k\nearrow1$, to obtain
\cref{ET3.2A} with $\varrho^*$ the subsequential limit
of $(1-\alpha)f^*_\alpha(\uptheta)$.
Concavity and local Lipschitz continuity is of course preserved at the limit.
\Cref{ET3.2B} follows from \cref{EL3.2A} and \cref{L3.2}\,(b),
while \cref{ET3.2C} follows directly from \cref{ET3.2A}.

Concerning part (b), \cref{ET3.2D} follows from \cref{ET3.2B}.
That the Foster--Lyapunov equation \cref{ET3.2E} implies
\cref{ET3.2F} is well-known from the theory of Markov processes
\cite[(V4)]{Meyn1993}.
Using \cref{ET3.2A,ET3.2F}, it is standard to show that any such selector
$q^*$ is optimal for the average cost criterion.
Thus in part (c), $\varrho^*$ is the optimal average cost
for the controlled Markov operator $\widehat\cT_q$ with running cost
$r_q$, while the term $\trace(\Pi^* DD\transp)$ arises from the plant.
\end{IEEEproof}

Since $\varrho^*$ is the optimal average cost, any subsequential limit
of $(1-\alpha)f^*_\alpha(\uptheta)$ must have this value.
Next we show uniqueness of solutions to \cref{ET3.2A} in a certain class
of functions.
This of course implies that any subsequential limit of \cref{PT3.2A}
leads to \cref{ET3.2A}.

We need the following lemma.
Recall the definitions in \cref{E-sets,E-hit}.

\medskip
\begin{lemma}\label{L3.3}
Suppose that $V$ is a nonnegative continuous function
which satisfies, for some positive constants $r$, $\theta>1$ and $\varepsilon\in(0,1)$,
\begin{equation}\label{EL3.3A}
\widehat\cT_{\Hat{q}} V(z) \,\le\, \varepsilon V(z)\quad
\forall\,z\in\cB_r^c\,,
\end{equation}
and
\begin{equation}\label{EL3.3B}
 \theta^{-1} \trace(\Pi^*\Sigma) \le V(s,\Sigma) \le \theta \trace(\Pi^*\Sigma)
\quad\forall (s,\Sigma)\in\SSp\times\cB_r^c\,.
\end{equation}
Then, for any $R>r$ we have
\begin{equation*}
\Exp^{\Hat{q}}_{z} \bigl[V(Z_{\uptau_R})\,\indic{\{\uptau_R<\Breve\uptau_r\}}\bigr]
\,\xrightarrow[]{R\to\infty}\,0\,\quad\forall z\in\cB_R\setminus\cB_r^c\,.
\end{equation*}
\end{lemma}

\begin{IEEEproof}
It is clear from \cref{EL3.3B}, that there exists a constant $m$, such
that
\begin{equation*}
\frac{V\bigl(s',\cT_{q}(\Sigma)\bigr)
+ V\bigl(s',\Xi(\Sigma)\bigr)}{V(s,\Sigma)}
\,\le\, m \quad \forall\,(s,\Sigma)\in\SSp\times\cB_r^c\,,\ \forall\,(s',q)\in\SSp\times\QSp\,.
\end{equation*}
Let $\delta \df \frac{\log\varepsilon}{2\log m}$.
It then follows that the function $V^{1+\delta}$ satisfies
$\widehat\cT_{\Hat{q}} V^{1+\delta}(z) \le \sqrt\varepsilon V^{1+\delta}(z)$
on $B_r^c$.
Therefore,
\begin{equation*}
\Exp^{\Hat{q}}_{z} \bigl[V^{1+\delta}(Z_{\uptau_R})\,
\indic{\{\uptau_R<\Breve\uptau_r\}}\bigr]\,\le\, V^{1+\delta}(z)\,.
\end{equation*}
Hence, we obtain
\begin{equation*}
\begin{aligned}
\Exp^{\Hat{q}}_{z} \bigl[V(Z_{\uptau_R})\,
\indic{\{\uptau_R<\Breve\uptau_r\}}\bigr]
&\,\le\, \Bigl(\sup_{\cB_R^c} V^{-\delta}\Bigr)\,
\Exp^{\Hat{q}}_{z} \bigl[V^{1+\delta}(Z_{\uptau_R})\,
\indic{\{\uptau_R<\Breve\uptau_r\}}\bigr]\\
&\,\le\, V^{1+\delta}(z)\, \sup_{\cB_R^c} V^{-\delta}
\,\xrightarrow[]{R\to\infty}\,0\
\end{aligned}
\end{equation*}
by \cref{EL3.3B}.
\end{IEEEproof}

We have the following result regarding the uniqueness of
the solution $f^*$ of \cref{ET3.2A}.

\medskip
\begin{theorem}\label{T3.3}
Any solution $f$ of \cref{ET3.2A} which is bounded from below
in $\cZ$ and has growth at most $\mathcal{O}\bigl(\trace(\Pi^*\Sigma)\bigr)$,
differs from $f^*$ by a constant.
\end{theorem}

\begin{IEEEproof}
By the proof of \cref{T3.2}, any such solution $f$ must satisfy
the lower bound in \cref{ET3.2C}.
Therefore, both $f$ and $f^*$ satisfy \cref{EL3.3A,EL3.3B},
for some $r>0$.
Let $q$ be a selector from the minimizer of the equation for $f$.
Then we have $\widehat\cT_q (f-f^*)\le f-f^*$ on $\cZ$,
and as a result we obtain
\begin{equation*}
\Exp_z^q\bigl[(f-f^*)(Z_{\Breve\uptau_r\wedge\uptau_R})\bigr]
\,\le\,(f-f^*)(z) \quad z\in \cB_R\setminus\cB_r^c\,,
\end{equation*}
which we write as
\begin{equation*}
\Exp_z^q\bigl[(f-f^*)(Z_{\Breve\uptau_r})\indic{\{\Breve\uptau_r\le\uptau_R\}}\bigr]
+\Exp_z^q\bigl[(f-f^*)(Z_{\Breve\uptau_R})\indic{\{\uptau_R<\Breve\uptau_r\}}\bigr]
\,\le\,(f-f^*)(z)\,,
\end{equation*}
and taking limits as $R\to\infty$, and using \cref{L3.3} and Fatou's lemma,
we obtain
\begin{equation*}
\inf_{\cB_r}\,(f-f^*) \,\le\,\Exp_z^q\bigl[(f-f^*)(Z_{\Breve\uptau_r})\bigr]
\,\le\,(f-f^*)(z)\quad\forall\,z\in \cB_r^c\,.
\end{equation*}
This shows that if we translate $f^*$ by an additive constant until it touches
$f$ at some point from below, it must touch it at some point in $\cB_r^c$.
Interchanging the role of $f$ and $f^*$, we see that a translate of $f$
also touches $f^*$ at some point from below in $\cB_r^c$.
This of course means that $f$ equals a translate of $f^*$, so they
differ by a constant.
\end{IEEEproof}

It is clear from \cref{T3.2} that the component $u^*$ of every optimal stationary
Markov control is as given in \cref{D3.2}.
Thus for the remainder of the paper we focus on the optimal querying problem,
or in other words, the optimal control problem in \cref{T3.2} with running
cost $c_q$.

\section{Relative Value Iteration}\label{S4}

The value iteration (VI) and relative value iteration (RVI)  algorithms 
generate a sequence of real-valued functions on $\SSp\times\psdef$ 
and associated constants 
that, as we show in \cref{T4.1}, converges to the
solution $f^*$ of  $(f^*,\varrho^*)$ of \cref{ET3.2A}.
Fix some point $\uptheta\in\cZ$. Without loss of generality, we may choose
the point $\uptheta$ in \cref{L3.2}.
Respectively, the VI and RVI are given by
\begin{align}
\varphi_{n+1} &\,=\, \min_{q\in\QSp}\,
\bigl\{c_q + \widehat\cT_{q}\,\varphi_{n}\bigr\} - \varrho^*\,,
 \label{E-VI}\tag{VI}\\
\rvi_{n+1}&\,=\,\min_{q\in\QSp}\,
\bigl\{c_q + \widehat\cT_{q}\,\rvi_{n}\bigr\}
 - \rvi_{n}(s_\circ,0)\,, 
 \quad
\rvi_0=\varphi_0\,,  \label{E-RVI}\tag{RVI}
\end{align}
where both algorithms are initialized with the same function
$\varphi_0\colon\cZ\to\RR_+$.
We let $\mathfrak{q}=\{\mathfrak{q}_m, m\in\NN\}$ denote a measurable
selector from the minimizer of \cref{E-VI}.  Note that since
the algorithms are initialized at the same $\varphi_0$, $\mathfrak{q}$ are
also minimizers for \cref{E-RVI} (and conversely).

We discuss some well-known properties of the algorithm.
First, it is clear that if $\varphi_0$ is continuous, concave, and non-decreasing, then
$\rvi_{n}$ and $\varphi_{n}$ are also 
continuous, concave, and non-decreasing for all $n > 0$.
The following identity is also easy to establish:
\begin{equation}\label{E-VI3}
\rvi_{n}(z)
 \,=\, \varphi_{n}(z)-\varphi_{n-1}(\uptheta) + \varrho^*\,.
\end{equation}

It follows from \cref{E-VI3} that if $\varphi_{n}$ converges pointwise to a function 
$\varphi^*\colon\SSp\times\psdef\to\RR$ as $n\to\infty$, 
then $\rvi_{n}$ converges to $\varphi^*-\varphi^*(\uptheta)+\varrho^*$.

With $c_n\equiv c_{q_n}$ and $\widehat\cT_n\equiv\widehat\cT_{q_n}$ we
write the VI as
$\varphi_{n+1}=\widehat\cT_n\varphi_{n}+ c_n -\varrho^*$.
Similarly, with $q_*$ denoting a selector from the minimizer of
\cref{ET3.2A}, and using the analogous definitions, we write the
optimality equation as
$f^* + \rho^* = c_* + \widehat\cT_*f^*$.
Since $\mathfrak{q}$ is suboptimal for \cref{ET3.2A},
and $q^*$ is suboptimal for \cref{E-VI}, it can be easily verified
that
\begin{equation}\label{E-VI4}
\widehat\cT_n(\varphi_{n}-f^*)
 \,\le\, \varphi_{n+1}-f^*
 \,\le\, \widehat\cT_*(\varphi_{n}-f^*)\,.
\end{equation}

Stability of the policies generated by the VI/RVI algorithms is usually not guaranteed,
nor is convergence of the iterates to the solution of the optimality equation
in \cref{ET3.2A}.
A popular control technique under the name of \emph{rolling horizon} control
consists of running a sufficiently large number of iterations
and computing the control $\mathfrak{q}_n\colon\cZ\to\QSp$ from the minimizer
of the $n^{\mathrm{th}}$ stage of the value iteration.
One would hope that, since $\mathfrak{q}_n$ is
the ``optimal action'' at the first step of the $n$-horizon problem,
then fixing a sufficiently large
$n$ and using $\mathfrak{q}_n$ as a stationary Markov policy
would result in a stable and near optimal control.
This is of course only a heuristic.
This problem is well
understood for finite state MDPs \cite{DellaVecchia2012} but it is 
decidedly unexplored for nonfinite state models.
Among the very few results in the literature is the study in \cite{Cavazos-Cadena1998}
for bounded running cost and under a simultaneous Doeblin hypothesis,
and the results in \cite{Hernandez-Lerma1990} under strong blanket stability
assumptions.
For the model considered here
there is no blanket stability; instead, 
the inf-compactness of the running cost 
penalizes unstable behavior.

In \cref{T4.1,T4.2} which follow we show that the rolling horizon control is
eventually stabilizing and its performance converges to the optimal one
as $n\to\infty$.
The bound in \cref{ET3.2D} plays a crucial role in this proof.
The same condition is used in \cite{AB-20}, but the convergence results cannot be applied
to our model in this paper.  The main reason is that the 
Markov process with kernel $\widehat\cT_q$ is not necessarily
irreducible under a stationary control.
Thus the usual supermartingale argument which establishes uniqueness of
the solution to \cref{ET3.2A} cannot be applied.
Thus we use a very different technique and show that
$\varphi_n\to f^* + \text{constant}$ for any solution $f^*$ of
\cref{ET3.2A} which satisfies \cref{ET3.2B}.
As a byproduct we establish uniqueness of solutions to \cref{ET3.2A} in
this class of functions.
The precise statement is as follows.

We first need a definition.
We say that a stationary Markov control $\Hat{q}\colon\cZ\to\QSp$ is geometrically
stable if there exists a function $\Lyap\colon\cZ\to\RR_+$,
constants $C_0$ and $\rho\in(0,1)$, and a compact set $K\subset\cZ$ satisfying
$\widehat\cT_{\Hat{q}}\Lyap \le C_0\indic{K} + \rho \Lyap$.

\begin{theorem}\label{T4.1}
Let $\rho=(1-\theta_1)$ and $\delta\df\frac{\rho^*+\theta_2}{\theta_1}$.
Suppose $\varphi_0$ is a nonnegative function in $\mathcal{O}(f^*)$.
Then the following hold:
\begin{itemize}
\item[(a)] 
There exists a constant $C_0$, such that $\varphi_n$ satisfies
\begin{equation}\label{ET4.1A}
(1-\rho^n) \bigl(f^*(z) -\delta\bigr) \,\le\, \varphi_n(z) 
\,\le\, f^*(z) + \bigl((\norm{\varphi_0}_{f^*}-1)\vee0\bigr)
\bigl(\delta+ \rho^n f^*(z)\bigr)\,,\quad\forall n\in\NN_0\,.
\end{equation}
\item[(b)]
The map $\mathfrak{q}_n\colon\cZ\to\QSp$ defines
a geometrically stable stationary Markov control
for any $n> \frac{\log\theta_1}{\log(1-\theta_1)}$.
\item[(c)] The compositions of $\widehat\cT_n$ remain tight for all $n> \frac{\log\theta_1}{\log(1-\theta_1)}$. In particular, we have
\begin{equation}\label{ET4.1B}
\begin{aligned}
\widehat\cT_n f^* &\,\le\, \frac{\rho}{1 - \rho^n}f^*
+ \delta \biggl(1+ \frac{1-\rho}{1- \rho^n}\biggr)\,,
\end{aligned}
\end{equation}
which implies that
\begin{equation}\label{ET4.1C}
\Exp^{{q_n}}_z \bigl[f^*(Z_t)\bigr] \,\le\, \delta\biggl(1 + \frac{1}{1-\rho - \rho^n}\biggr)
+ \biggl(\frac{\rho}{1 - \rho^n}\biggr)^t f^*(z)\quad\forall\,z\in\cZ\,,\ \forall\,n > \frac{\log\theta_1}{\log(1-\theta_1)} \,.
\end{equation}
\end{itemize}
\end{theorem}

\begin{IEEEproof}
The lower bound in \cref{ET4.1A} is identical to \cite[Theorem~6.1]{AB-20}.
The upper bound is obtained from \cref{E-VI4} using
\cref{ET3.2F}.

We continue with part (b).
Without loss of generality we assume that $\varphi_0=0$.
By \cref{ET3.2D,ET4.1A}, we obtain
\begin{equation}\label{PT4.1A}
\begin{aligned}
\abs{\varphi_{n+1}-\varphi_n} &\,\le\,
(1-\rho^n) \delta + \rho^n \,f^*(z) \\
&\,\le\,
(1-\rho^n) \delta
+\tfrac{\theta_2}{\theta_1}\rho^n + \tfrac{\rho^n}{\theta_1} c_n\,.
\end{aligned}
\end{equation}
Therefore, we obtain from \cref{E-VI,PT4.1A}, that
\begin{equation}\label{E-poiss}
\begin{aligned}
\widehat\cT_n \varphi_n &\,=\, -(c_n - \varrho^* -\varphi_{n+1}+\varphi_n) + \varphi_n \\
&\,\le\, (1-\rho^n) \delta
+\tfrac{\theta_2}{\theta_1}\rho^n + \varrho^*
-\bigl(1-\tfrac{\rho^n}{\theta_1}\bigr)c_n+ \varphi_n\,.
\end{aligned}
\end{equation}
It follows that $\mathfrak{q}$ is geometrically stable
if $\tfrac{\rho^n}{\theta_1}<1$ which proves the result. In addition,
\begin{equation}\label{PT4.1B}
\begin{aligned}
\widehat\cT_n f^* &\,=\,
\widehat\cT_n \varphi_n + \widehat\cT_n (f^*-\varphi_n)\\
&\,=\, \varphi_{n+1} -c_n +\varrho^* + \widehat\cT_n (f^*-\varphi_n)\\
&\,=\, (f^* -c_n) +\varrho^* + \widehat\cT_n (f^*-\varphi_n) + (\varphi_{n+1} - f^*)\\
&\,\le\, \rho f^* + \theta_2 + \varrho^* + \widehat\cT_n (f^*-\varphi_n)\\
&\,\le\,\rho f^* + \theta_2 + \varrho^* + (1-\rho^n)\delta + \rho^n \widehat\cT_n f^*\,,
\end{aligned}
\end{equation}
where the first inequality follows from \cref{ET3.2D,ET4.1A} and the second by
applying $\widehat\cT_n$ to \cref{ET4.1A}.
Note that $\frac{\rho}{1- \rho^n}<1$ for any
$n > n_0  = \frac{\log\theta_1}{\log(1-\theta_1)}$.
Therefore, \cref{PT4.1B} implies \cref{ET4.1B} which in its turn implies \cref{ET4.1C}.
The compositions of $\widehat\cT_n$ being tight is a direct consequence of \cref{ET4.1B}.
\end{IEEEproof}

We need the following assumption.

\begin{assumption}\label{A4.1}
Let $$G(z) \df \{z'\in\cZ\colon \Prob^{q^*} (Z_{t+1}=z'\,|\, Z_t=z)>0\}\,,$$
and $G^n$ be the $n^{\mathrm{th}}$ composition of this set valued map.
In other words $G^n$ is the support of the $n^{\mathrm{th}}$ composition
of $\Prob^{q^*}_z$, or otherwise stated, the endpoints of all
optimal sample paths of length $n$ starting at $z$.
For each $z,z'\in\cZ$, there exists $r>0$ such that
\begin{equation*}
\dist\Bigl(\cB_r\cap\cup_{n\in\NN} G^n\bigl(\xi(z)\bigr),\cB_r\cap\cup_{n\in\NN} G^n\bigl(\xi(z')\bigr)\Bigr) =0\,,
\end{equation*}
where $\dist$ is a metric on $\cZ$ and $\xi(z) \df (s, \widetilde{\Pi}^* \Sigma)$ for $z = (s,\Sigma)$.
\end{assumption}

\begin{theorem}\label{T4.2}
Under \cref{A4.1} the following hold:
\begin{itemize}
\item[(a)]
As $n\to\infty$, $\varphi_{n}-f^*$ converges to a constant,
and $\widetilde\varphi_n$ converges to 
$f^*-f^*(\uptheta)+\varrho^*$.
\item[(b)]
The stationary Markov control $\mathfrak{q}_n$ is asymptotically optimal in
the sense that $J^{\mathfrak{q}_n}\to J^*$ as $n\to\infty$.
\end{itemize}
\end{theorem}


\begin{IEEEproof}
Let $\varphi_0$ be a continuous and concave initial condition
in $\mathcal{O}\bigl(f^*)$.
By \cref{T4.1} the orbit $\{\varphi_t\}_{t\in\NN_0}$
under the semigroup of transformations defined by \cref{E-VI} is
precompact and its limit points, that is, the $\omega$-limit set
$\omega(\varphi_0)$ is a compact set which satisfies the bound
\begin{equation}\label{PT4.2A}
-\delta \,\le\, h - f^* \,\le\, \bigl((\norm{\varphi_0}_{f^*}-1)\vee0\bigr)\delta
\end{equation}
for all $h\in\omega(\varphi_0)$.
It follows then from the right-hand side inequality of
\cref{PT4.2A} that the quantity
$\overline{m} =\overline{m}(h) \df \sup_{\cZ}\,(h-f^*)$ 
must be a constant independent of $h\in\omega(\varphi_0)$.
Without loss of generality, and using invariance under translations
of the solutions to \cref{E-VI},
we can assume $\overline{m}=0$.
Thus $h\le f^*$ for all $h\in\omega(\varphi_0)$.

Consider an orbit $\{h_n\}_{n\in\NN_0}\subset\omega(\varphi_0)$ under
the VI.
By \cref{E-VI4}, $\{(h_{n-k}-f^*)(Z_k)\}_{k=0,\dotsc,n}$ is a submartingale. Using the hitting time
$\Breve\uptau_r\wedge n$ and optional sampling, we therefore have
\begin{equation}\label{PT4.2B}
(h_n-f^*)(z) \,\le\,
\Exp^*_z\bigl[(h_{n-\Breve\uptau_r\wedge n}-f^*)(Z_{\Breve\uptau_r\wedge n})\bigr]\,.
\end{equation}

For an arbitrary $\epsilon$,
select any $g\in \omega(\varphi_0)$, and a point $z$ such that
$g(z)-f^*(z)\ge -\epsilon$.  This is possible by the hypothesis that $\overline{m}=0$.
By the invariance of $\omega(\varphi_0)$ there exists a sequence $n_k$
such that $h_{n_k}\to g$ as $n_k\to\infty$, and therefore
$h_{n_k} - g\ge -2\epsilon$ for all but finite terms of this sequence.
Note that $\Exp^*_z[V(Z_{\Breve\uptau_r})\indic{\{\Breve\uptau_r>n\}}]\to0$
as $n\to\infty$ for any function $V\in\mathcal{O}(f^*)$.
Hence, evaluating \cref{PT4.2B} at $n=n_k$, and then sending $k\to\infty$, we obtain
\begin{equation}\label{PT4.2C}
-3\epsilon \,\le\,
\Exp^*_z\bigl[(h_{n_k-\Breve\uptau_r}-f^*)(Z_{\Breve\uptau_r})
\indic{\{\Breve\uptau_r\le n_k\}}\bigr]\,,
\end{equation}
for all $n_k>\Bar{n}$ for some $\Bar{n}$.
Since both $\epsilon>0$ and $g$ are arbitrary, \cref{PT4.2C} shows the following:
For every $h\in\omega(\varphi_0)$ there exists $z\in\cB_r$ such that
$h(z)= f^* + \overline{m}$.

Next we study the $\Argmin (h-f^*)$
using  the left-hand side inequality of \cref{E-VI4}.
Shift $h$ so that $\inf (h-f^*)=0$ and consider
a sequence $t_n$ such that $h_{t_n}\to h$.
Such a sequence exists by invariance of $\omega(\varphi_0)$.
Extract a further subsequence so that $t_{n+1}-t_n\to\infty$.
Suppose that $h(z) - f^*(z)\le\epsilon$.
Then the lhs of \cref{E-VI4} gives
\begin{equation}\label{PT4.2D}
\Exp^{q_n}_z\bigl[(h_{t_n} - f^*)(Z_{t_{n+1}-t_n})\bigr]
\,\le\, (h_{t_{n+1}} - f^*)(z) \,.
\end{equation}
The rhs of \cref{PT4.2D} converges to $\le\epsilon$.
Recall from \cref{ET4.1B} that the compositions of $\widehat{\cT}_n$ are tight. Choose a compact set $\cB_r$ that has at least half
the total mass of these compositions.
This means that $\inf_{\cB_r} (h-f^*)\le 2\epsilon$ (plain Markov inequality).
Since $\epsilon$ is arbitrary, the infimum of $h-f^*$, that is $0$,
must be realized on $\cB_r$.
Let $z\in\cB_r$ be such a point, and consider \cref{E-VI4}
with $h_1\equiv h$.
Since $h_1(z)=f^*(z)$ and $h\ge f^*$ for all $h\in\omega(\varphi_0)$, using the monotonicity of the dynamic
programming operator, we obtain 
\begin{equation}\label{PT4.2E}
h_1 +\rho^* = c_0 + P_0 h_0 = c_* + P_* h_{0}\,.
\end{equation}
So at points where the minimum (or the maximum) of the difference
of $h_1-f^*$ is attained the minimizer in \cref{E-VI} is the stationary $q^*$.

Let $z=z_{n+1}\in\Argmin(h_{t_{n+1}} - f^*)$.
Since $z_{n+1}$ attains the minimum in $h_{t_{n+1}} - f^*$,  all the subsequent
iterates until we hit $h_{t_n}$ on the lhs of \cref{PT4.2D} must attain it also.
And by the observation we made in \cref{PT4.2E}, the expectation $\Exp^{q_n}$ in
\cref{PT4.2D} is the same as $\Exp^*$.
Thus taking limits, and since $h_{t_n}\to h$, the following holds:
For every $h\in\omega(\varphi_0)$ there exists $z'\in\cB_{r'}$ such that
$h(z')= f^*$.

In the above,
 a sequence of concave functions $h_k$, $k\in \NN_0$,
 with $h_k \in \omega(\varphi_0)$
 was constructed such that $h_{n+1} = \min_{q\in\QSp}\, \bigl\{c_q + \widehat\cT_{q}\,h_{n}\bigr\} - \varrho^*  $, $\sup_{\cZ}\,(h-f^*) = \overline{m}>0$, and $\inf_{\cZ}\,(h-f^*) = 0$. In addition, both the supremum and the infimum of the difference of $h_n-f^*$ are attained at $z$ and $z'$ respectively and the minimizer at these points is the stationary $q^*$. 
Note also that $\widehat\cT^k_{*}(h_n - f^*)(z) = \overline{m}$ and $\widehat\cT^k_{*}(h_n - f^*)(z') = 0$ where $\widehat\cT^k_{*}(h_n - f^*)(z)$ denotes the $k^{th}$ composition of
$\widehat\cT_{*}$ of $h_n - f^*$ starting at $z$.

Using \cref{T3.2,T3.3}, recall that $f^*$ is the unique (up to a constant) Lipschitz continuous solution of \cref{ET3.2A}. Recall also that $\varphi_0 \in \mathcal{O}(f^*)$ which means $(h_n - f^*) \in \mathcal{O}(f^*)$ and is a Lipschitz continuous function using \cref{L2.1} with a positive Lipschitz constant $\kappa$ for all k. Under \cref{A4.1}, there exists $r>0$ such that the union of the supports of $(\Prob_z^{q^*})^{k}$ starting at $z$ and $z'$ are at a 0 distance inside the ball $\cB_r$ where $(\Prob_z^{q^*})^{k}$ denotes the $k^{th}$ composition of $\Prob_z^{q^*}$. This contradicts the  Lipschitz continuity of $h_n - f^*$ and concludes the proof of part (a). 

To prove part (b), rewrite \cref{E-poiss} as $
\widehat\cT_n \varphi_n \,=\, g_n + \varphi_n\,,$
where $g_n = -(c_n - \varrho^* -\varphi_{n+1}+\varphi_n)$. Iterating this sequence to obtain
\begin{equation*}
\frac{1}{N} \widehat\cT_n^N \varphi_n = \frac{1}{N} \sum_{i = 0}^{N-1}\widehat\cT_n^i g_n + \frac{1}{N}\varphi_n\,, 
\end{equation*}
where $\widehat\cT_n^i$ denotes the $i^{th}$ composition of $\widehat\cT_n$. Using \cref{ET4.1A,ET4.1B}, one can conclude that 
\begin{equation*}
\lim_{N \to \infty} \frac{1}{N} \widehat\cT_n^N \varphi_n \,=\,\lim_{N \to \infty} \frac{1}{N} \varphi_n  \,=\, 0\,.
\end{equation*}

Since $\varphi_{n+1} - \varphi_n \in \mathcal{O}(c_n)$ and by part (a) we have $\varphi_{n+1} - \varphi_n \to 0$ as $n \to \infty$, we have
\begin{equation*}
\lim_{n \to \infty} \lim_{N \to \infty}\frac{1}{N} \sum_{i = 0}^{N-1}\widehat\cT_n^i c_n = \varrho^*\,.
\end{equation*}
This completes the proof.
\end{IEEEproof}

\section{Sensor-Dependent Loss Rates}\label{S5}
We now turn our attention to a special case of the previous results, 
with a single network state. In this case, the network cost is simply
a function of the query process $\{Q_{t}\}$, taking values in 
the finite set of allowable sensor queries $\QSp$. 
The loss rate depends only on the query, so
it can be treated as a vector $\lambda$ in $[0,1]^{\abs{\QSp}}$,
indexed by the corresponding query:
\begin{equation}\label{E-loss2}
\Prob(\gamma=1) \,=\, (1-\lambda_{q})\,, \qquad
\Prob(\gamma=0) \,=\, \lambda_{q}\,,
\end{equation}
for $q\in\QSp$.
We are interested in characterizing the set 
of loss rates $\Lambda_{s}\subset [0,1]^{\abs{\QSp}}$
for which the system is stabilizable. 
Our formulation generalizes the problem in \cite{Sinopoli2004},
which analyzes the system \cref{E-LQG}--\cref{E-LQGY} without
sensor scheduling ($C_{q}=C$) and therefore with a single loss rate.

The authors in \cite{Sinopoli2004} prove
that there is a critical loss rate $\lambda_{c}\in (0,1)$ such that
the system is stabilizable if and only if $\Bar{\lambda} < \lambda_{c}$
(i.e., $\Lambda_{s}=[0,\lambda_{c}$).
Here, we generalize that result, showing that when selecting different 
sensors induces different loss rates, there is a critical surface 
$\mathcal{W}\subset [0,1]^{\abs{\QSp}}$.
The system is stabilizable if and only if the vector
$\lambda < \lambda'\in\mathcal{W}$.

Recalling the discussion around \cref{A3.1},
$\Lambda_{s}=\varnothing$ unless $(A,B)$ is stabilizable
and $(\overline{C},A)$ is detectable. 
Hence, without loss of generality, we assume 
$(A,B)$ is stabilizable and $(\overline{C},A)$ is detectable
and therefore, by the results in \cite{Wu2008}, $0\in\Lambda_{s}$.

In order to distinguish between operations with different loss rates,
we will indicate the corresponding rate in a superscript, as in 
\begin{equation*}
\widehat\cT_{q}^{\lambda}f(\Sigma)
=(1-\lambda_{q})f(\cT_{q}(\Sigma))+\lambda_{q}f\bigl(\Xi(\Sigma)\bigr)\,.
\end{equation*}
We start with the following theorem.

\begin{theorem}\label{T5.1}
If the system \cref{E-LQG}--\cref{E-LQGY} with \cref{E-loss2} 
is stabilizable for a loss rate $\lambda'\in [0,1]^{\abs{\QSp}}$, 
then it is also stabilizable for 
any other loss rate $\lambda\le\lambda'$.
In other words, the set $\Lambda_{s}$ is order-convex with respect
to the natural ordering of positive vectors in $\RR^{\abs{\QSp}}$.
\end{theorem}
\begin{IEEEproof}
Suppose that $\lambda'\in \Lambda_{s}$, 
and let $\{Q_{t}\}$ be a stabilizing query sequence.
Let $\lambda\in [0,1]^{\abs{\QSp}}$ such that $\lambda\le\lambda'$.
For a non-decreasing function $f\colon\psdef\to\RR$ and any $q\in\QSp$,
\begin{equation*}
\widehat\cT_{q}^{\lambda}f(\cdot) - \widehat\cT_{q}^{\lambda'}f(\cdot)
\,=\, (\lambda'_{q}-\lambda_{q})
\bigl(f\bigl(\cT_{q}(\cdot)\bigr)-f\bigl(\Xi(\cdot)\bigr)\bigl) \,\le\, 0\,.
\end{equation*}
Applying to $\trace(\cdot)$, which is non-decreasing in $\psdef$, we get
\begin{equation}\label{PT5.1A}
\widehat\cT_{q}^{\lambda}\trace(\cdot)
 - \widehat\cT_{q}^{\lambda'}\trace(\cdot)
= - (\lambda'_{q}-\lambda_{q})\,
\trace\bigl(\widehat{K}_{q,1}(\cdot)C_{q}\Xi(\cdot)\bigr) 
\,\le\, 0\,,
\end{equation}
since $\widehat{K}_{q,1}(\Sigma)C_{q}\Xi(\Sigma)\in\psdef$.
Iterating \cref{PT5.1A} with the stabilizing
query sequence yields
\begin{equation*}
\Exp_{\Sigma_0}^{Q_{t},\lambda}\bigl[\trace(\widetilde \Pi^*\widehat\Pi_t)\bigr]
\le
\Exp_{\Sigma_0}^{Q_{t},\lambda'}\bigl[\trace(\widetilde \Pi^*\widehat\Pi_t)\bigr],
\quad \text{ for all } t\ge 0,
\end{equation*}
and stability with $\lambda$ follows.
\end{IEEEproof}

Moreover, a lower loss rate leads to a smaller error covariance
at every time step.
We continue with another important result.

\begin{theorem}\label{T5.2}
If the system \cref{E-LQG}--\cref{E-LQGY} with \cref{E-loss2} 
is stabilizable for a loss rate $\lambda\in [0,1]^{\abs{\QSp}}$, 
there exists an open neighborhood $\cB\subset [0,1]^{\abs{\QSp}}$
around $\lambda$ such that the system is stabilizable for $\lambda'\in\cB$.
\end{theorem}
\begin{IEEEproof}
Let $\lambda\in \Lambda_{s}$, and let $f^*$ be the solution
of \cref{ET3.2A} with a loss rate $\lambda$ and $q^*$ a selector of the minimizer. Let $\kappa_1$ be a constant such that $\kappa_1 \Sigma - A \Sigma A\transp \ge 0$ for any $\Sigma \in \psdef$. 
Recalling the constants from \cref{ET3.2B}, 
let $\lambda'\ge\lambda$ such that for a particular $\Bar{q}\in\QSp$,
\begin{equation*}
0 < \lambda'_{\Bar{q}}-\lambda_{\Bar{q}}<
\frac{1}{\kappa_0^* \kappa_1} \,, 
\end{equation*}
and $\lambda'_{q}=\lambda_{q}$ for all $q\in\QSp\setminus\{\Bar q\}$.
Then, using the bound \cref{ET3.2B},
\begin{equation*}
\begin{aligned}
\widehat\cT_{\Bar{q}}^{\lambda'}f^*(\Sigma) - f^*(\Sigma)
&\,=\,
(\lambda'_{\Bar{q}}-\lambda_{\Bar{q}})
\bigl(f^*(\Xi(\Sigma))-f^*(\cT_{\Bar{q}}(\Sigma))\bigr)
-\trace(\widetilde\Pi^*\Sigma)+\varrho^*-\rcnt(\Bar{q}) \\
&\,\le\, (\lambda'_{\Bar{q}}-\lambda_{\Bar{q}})
f^*(\Xi(\Sigma)) -\trace(\widetilde\Pi^*\Sigma) +\varrho^*-\rcnt(\Bar{q}) \\
&\,\le\, \kappa_0^* (\lambda'_{\Bar{q}}-\lambda_{\Bar{q}})
\Bigl(1 + \trace \bigl(\widetilde\Pi^* \Xi(\Sigma)\bigr)\Bigr)
-\trace(\widetilde\Pi^*\Sigma)+\varrho^*-\rcnt(\Bar{q}) \\
&\,\le\, \bigl(\kappa_0^* \kappa_1 (\lambda'_{\Bar{q}}-\lambda_{\Bar{q}}) 
 - 1 \bigr) \trace(\widetilde\Pi^*\Sigma) \\
&\mspace{230mu}+ \kappa_0^* (\lambda'_{\Bar{q}}{-}\lambda_{\Bar{q}})\bigl(\trace(\widetilde\Pi^*DD\transp)+ 1\bigr)
+\varrho^*{-}\rcnt(\Bar{q}) \\
&\,\le\, 
-\frac{1- \kappa_0^* \kappa_1 (\lambda'_{\Bar{q}}-\lambda_{\Bar{q}})}{\kappa_0^*} \,f^*(\Sigma) + \Bar{M}
\end{aligned}
\end{equation*}
for some $\Bar{M}\in\RR$. 
Hence the chain with rate $\lambda'$ is stable under 
$\bar q$. 
Applying this result and \cref{T5.1} to each index,
we find that a loss rate vector $\Bar{\lambda}$ is in $\Lambda_{s}$ if
\begin{equation*}
(\Bar{\lambda}_{q}-\lambda_{q})^+ <
\frac{1}{\kappa_0^* \kappa_1}
\quad \forall q\in\QSp\,.
\end{equation*}
This completes the proof.
\end{IEEEproof}

Combining these results, we obtain the following corollary concerning the 
structure of $\Lambda_{s}$.

\begin{corollary}
Suppose that $(A,B)$ is stabilizable and $(\overline C, A)$ is detectable.
Then, there exists a critical surface $\mathcal{W}$ in $(0,1]^{\abs{\QSp}}$ such that 
the system is stabilizable with loss rate $\lambda$ 
if and only if $\lambda < \lambda'\in\mathcal{W}$.
More precisely, there exists a function 
$\mathscr{F}\,\colon\RR^{\abs{\QSp}-1}\to[0,1]$
which is nonincreasing in each argument such that
the system is stabilizable with loss rate $\lambda$ if and only
if
$\lambda_{\abs{\QSp}}<
\mathscr{F}(\lambda_1,\dotsc,\lambda_{\abs{\QSp}-1})$.
In other words, $\Lambda_{s}$ is the strict hypograph of $\mathscr{F}$.
\end{corollary}

\begin{definition}
A set of sensor queries $\QSp$ is called \textit{non-redundant} 
if the system is not stabilizable
without querying each sensor infinitely often.
That is, the system is stabilizable if $\lim_{T \to \infty}\frac{1}{T}\Exp_{z}
 \bigl[\sum_{t=0}^{T-1} \indic{Q_{t} = q}\bigr] >0$ for all $q \in \QSp$
and $z \in \cZ$.
\end{definition}

\begin{theorem}\label{T5.3}
Suppose that the set of sensors is non-redundant and that 
$\lambda,\lambda'\in \Lambda_{s}$ such that $\lambda \le \lambda'$
and $\lambda\neq\lambda'$. Then $\varrho^*_{\lambda}\, <\, \varrho^*_{\lambda'}$.
\end{theorem}

\begin{IEEEproof}
Without loss of generality, let $\lambda,\lambda'\in \Lambda_{s}$ such that 
$\lambda_{\Bar{q}}<\lambda_{\Bar{q}}'$ and
and $\lambda_{q}=\lambda'_{q}$ for all $q\in\QSp\setminus\{\Bar{q}\}$.
For the system with loss rate $\lambda$ (respectively, $\lambda'$), 
let $f_{\lambda}^*$ ($f_{\lambda'}^*$) be the solution of \cref{ET3.2A}, 
and let $q^{\lambda}$ ($q^{\lambda'}$) be a selector of the corresponding minimizer.
The function $f_{\lambda'}^*$ is non-decreasing, so
for any query $q\in\QSp$ we have
\begin{equation*}
\widehat\cT^{\lambda'}_{q}f_{\lambda'}^*(\cdot)
-\widehat\cT^{\lambda}_{q}f_{\lambda'}^*(\cdot)
\,=\, (\lambda_{q}'-\lambda_{q})
 \bigl(f_{\lambda'}^*(\Xi(\cdot)) {-} f_{\lambda'}^*(T_{q}(\cdot))\bigr) \,\ge\,0 \,.
\end{equation*}
Since $\QSp$ is non-redundant and $f^*$ is a concave function such that
$f^*(\Sigma) \ge \trace (\widetilde \Pi^* \Sigma)$, we have
\begin{equation}\label{PT5.3A}
\lim_{T \to \infty}\frac{1}{T}\Exp_{\Sigma}^{\lambda,q^{\lambda'}}
 \Biggl[\sum_{t=0}^{T-1} \Bigl(f_{\lambda'}^*\bigl(\Xi((\widehat\Pi_{t}))\bigr) - f_{\lambda'}^*\bigl(T_{\Bar{q}}((\widehat\Pi_{t}))\bigr)\Bigr) \indic{Q_{t} = \Bar{q}} \Biggr]\,>\,0\,,
\end{equation}
for all $\Sigma \in \psdef$.
If not, then a different sensor
could be queried instead of $\Bar q$
and the system would still be stable.

Define the non-negative function 
$g_{q}(\Sigma)\df \widehat\cT^{\lambda'}_{q}f_{\lambda'}^*(\Sigma)
-\widehat\cT^{\lambda}_{q}f_{\lambda'}^*(\Sigma)$.
Then, for any $\Sigma\in\psdef$, 
\begin{equation*}
\begin{aligned}
\varrho^*_{\lambda'}
&\,=\, \rcnt(q^{\lambda'}(\Sigma)) + \trace(\widetilde\Pi^* \Sigma)
 +\widehat\cT^{\lambda'}_{q^{\lambda'}}f_{\lambda'}^*(\Sigma)
 - f_{\lambda'}^*(\Sigma) \\
&\,=\, \rcnt(q^{\lambda'}(\Sigma)) + \trace(\widetilde\Pi^* \Sigma)
 + g_{\Bar{q}}(\Sigma)\indic{q^{\lambda'}(\Sigma) = \Bar{q}}
 +\widehat\cT^{\lambda}_{q^{\lambda'}}f_{\lambda'}^*(\Sigma)
 - f_{\lambda'}^*(\Sigma) \\
&\,=\, \frac{1}{T}\Exp_{\Sigma}^{\lambda,q^{\lambda'}}
 \Biggl[\sum_{t=0}^{T-1} \rcnt(Q_{t}) + \trace(\widetilde\Pi^* \widehat\Pi_t) \Biggr] 
+ \frac{1}{T}\Exp_{\Sigma}^{\lambda,q^{\lambda'}}
 \Biggl[\sum_{t=0}^{T-1} g_{Q_{t}}(\widehat\Pi_{t}) \indic{Q_{t} = \Bar{q}} \Biggr]\\
&\mspace{300mu}+\frac{1}{T}\Exp_{\Sigma}^{\lambda,q^{\lambda'}}
\bigl[f_{\lambda'}^*(\widehat\Pi_{T}) - f_{\lambda'}^*(\widehat\Pi_0)\bigr]\,.
\end{aligned}
\end{equation*}  
Taking limits as $T\to\infty$, the third term on the right hand side
approaches $0$, and using \cref{PT5.3A},
this shows that
$\varrho_{\lambda'}^* >J_{\lambda}^{q^{\lambda'}}$,
where $J_{\lambda}^{q^{\lambda'}}$ is the average cost for the system 
with loss rate $\lambda$ and using policy $q^{\lambda'}$.
Since $q^{\lambda'}$ is suboptimal, it follows that
$\varrho^*_{\lambda}\le J_{\lambda}^{q^{\lambda'}} < \varrho^*_{\lambda'}$.
\end{IEEEproof}

\begin{remark}
Since the average cost $\varrho^*_{\lambda}\to\infty$
as the system parameters approach the boundary of
the stability region, the set 
$\Lambda(\kappa)\df \{\lambda \colon\varrho_{\lambda}^*<\kappa\}$ is a 
ray-connected neighborhood of $0$ for all $\kappa>0$. 
Clearly, $\bigcup_{\kappa>0}\Lambda(\kappa)=\Lambda_{s}$.
\end{remark}

\begin{remark}
Note that similar results could be shown for the more general case
with network states dictating loss rates. However, the analysis
is much more involved, and may require additional assumptions on the 
structure of the network state transition probabilities. We 
present the simpler version here to facilitate the analysis and the comparison 
to the previous works.
\end{remark}

\begin{remark}
Suppose that the loss rates depend only on the query, as in \cref{E-loss2}, but are
unknown.
Then the implications of Theorem \ref{T5.2} are remarkable. 
Since stability is shown to be an open property, if one can
find an estimator sequence $\widehat{\lambda}_{t}\to\lambda$ a.s., then
the system will retain stability and the long-term average performance
would be the same as if the rates were known beforehand. 
Since the channel is Bernoulli, recursive estimation of the loss rates
leading to a.s.\ convergence to the true value is rather straightforward.
For example, a maximum likelihood estimator can be employed, 
as in \cite{Fernandez-Gaucherand1993}.
\end{remark}

\subsection{Diagonal Structures}

Without loss of generality we focus on the estimation problem.
Consider two one-dimensional systems
\begin{equation}\label{E-diag}
\begin{split}
x^{(i)}_{k+1} &\,=\, a_{i} x^{(i)}_{k} + w^{(i)}_{k}\\[5pt]
y^{(i)}_{k} &\,=\, x^{(i)}_{k} + f_{i} v^{(i)}_{k}\,,
\end{split}
\end{equation}
where $\{w^{(i)}_{k}, v^{(i)}_{k}\,, k\in\NN\,,\;i=1,2\}$ are i.i.d. standard
Normal random variables.
Note that we can always scale the system so that $c_{i}=1$ and $w^{(i)}_{k}$
has unit variance, so the above representation is without loss of generality.
It is well known that the Kalman filter with intermittent observations
is stable for each subsystem separately
if and only if $\lambda_{i}<(\max_i a_i)^{-2}$ \cite{Sinopoli2004}.

We concentrate on the case where $a_1=a_2=a$.
We assume that $a>1$, otherwise the problem is trivial.
Suppose that the intermittency rate is of the form
$(\lambda,\lambda)$ with $\lambda\in[0,a^{-2})$.
Let $\xi_{1}$ and $\xi_{2}$ be the estimation error variances
of $x^{(1)}$ and $x^{(2)}$, respectively, and define
$\xi\df(\xi_{1},\xi_{2})$.
Note that
\begin{equation*}
\cT_{1}(\xi) \,=\, \biggl(\frac{f_{1}^{2}\,(1+a^{2}\xi_{1}) }
{1+a^{2}\xi_{1}+ f_{1}^{2}}, 1+a^{2}\xi_{2}\biggr)\,,
\end{equation*}
and the analogous expression holds for $\cT_{2}$.
We have the bound
\begin{equation*}
\frac{f_{i}^{2}\,(1+a^{2}\zeta)}
{1+a^{2}\zeta+ f_{i}^{2}}\,\le\,\max\,(f_1^2,f_2^2)\qquad\forall\, \zeta\in\RR_{+}\,,
\quad i=1,2\,.
\end{equation*}
For $\varepsilon>0$,
let $\Lyap_{\varepsilon}\colon \RR^{2}_{+}\to\RR_{+}$ be defined as follows.
\begin{equation*}
\Lyap_{\varepsilon}(\xi)\,\df\,
\begin{cases}
\varepsilon\,\xi_{1}+ (1-\varepsilon)\,\xi_{2}\,,&\text{if~} \xi_{1}\ge\xi_{2}\\[3pt]
(1-\varepsilon)\,\xi_{1}+ \varepsilon\,\xi_{2}\,,&\text{otherwise.}
\end{cases}
\end{equation*}
Let $\varepsilon>0$ be small enough such that
\begin{equation}\label{E-eps}
\varepsilon_{0}\,\df\,\bigl(\tfrac{\varepsilon}{1-\varepsilon}+\lambda\bigr) a^2\,<\,1\,.
\end{equation}
Suppose $m_{0}\df \max\,(f_1^2,f_2^2)\le\xi_{2}\le \xi_{1}$.
We have 
\begin{align*}
\widehat\cT_{1}^{\lambda}\Lyap_{\varepsilon}(\xi) - \Lyap_{\varepsilon}(\xi)&\,=\,
(1-\lambda) \Lyap_{\varepsilon} \bigl(\cT_{1}(\xi)\bigr)
+\lambda \Lyap_{\varepsilon}
\bigl(1+a^{2}\xi_1,1+a^{2}\xi_2\bigr) - \Lyap_{\varepsilon} (\xi)\\[2pt]
&\,\le\,
(1-\lambda)\bigl[(1-\varepsilon)\,m_{0} + \varepsilon\,(1+a^{2}\xi_{2})\bigr]
\\&\mspace{50mu}
+\lambda\bigl[\varepsilon\,(1+a^{2}\xi_{1})  + (1-\varepsilon)\,(1+a^{2}\xi_{2})\bigr]
- \Lyap_{\varepsilon} (\xi)\\
&\,\le\, C_{0} +\Bigl((1-\lambda)\tfrac{\varepsilon\,a^2}{1-\varepsilon}
+\lambda\,a^{2} -1\Bigr) \Lyap_{\varepsilon} (\xi)\\
&\,\le\, C_{0} - (1-\varepsilon_{0}) \Lyap_{\varepsilon} (\xi)\,,
\end{align*}
where $C_{0}$ is a constant depending on $\varepsilon$, $\lambda$, and $m_{0}$.
On the other hand, if $\xi_{1}\ge\xi_{2}$, and $\xi_{2}<m_{0}$,
then $\Lyap_{\varepsilon} \bigl(\cT_{1}(\xi)\bigr)$ is bounded and we obtain
\begin{equation*}
\widehat\cT_{1}^{\lambda}\Lyap_{\varepsilon}(\xi) - \Lyap_{\varepsilon}(\xi)
\,\le\, C_{0}' + (\lambda a^{2}-1) \Lyap_{\varepsilon} (\xi)
\end{equation*}
for some constant $C'_0$.
Therefore, by symmetry, we obtain
\begin{equation*}
\min_{q=1,2}\,\widehat\cT_{q}^{\lambda}\Lyap_{\varepsilon}(\xi) - \Lyap_{\varepsilon}(\xi)
\,\le\, C''_{0} - (1-\varepsilon_{0}) \Lyap_{\varepsilon} (\xi)\qquad
\forall\, \xi\in\RR_{+}^2
\end{equation*}
for some constant $C''_{0}$.
Since $(1-\varepsilon_{0})>0$ by \cref{E-eps}, stability follows.
The same technique applies for a diagonal system as in \cref{E-diag}
of any order.
Thus we have proved the following.

\begin{theorem}
Consider a system in diagonal form as in \cref{E-diag}, with
$a_{i}=a>1$, $i=1,\dotsc,N_q$.
Then $\Lambda_{s} \,=\, [0,\nicefrac{1}{a^{2}})^{N_q}$.
\end{theorem}

%

\section{Conclusion}\label{sec-conclusion}
This paper presents a new stability result for optimal sensor scheduling with a more general and
flexible structure than previous results. 
The inclusion of network dynamics allows for a variety of interesting and practical applications
of distributed sensing and control. 
Further, we show that the results are not simply theoretical. 
The convergence of the value iteration algorithm, combined with the approximations 
on bounded subsets of the state space and rolling horizon policies, implies that
practical results and applications of the theory are achievable. 
This richness of results is surprising given the generality of the system assumptions, as previous
results of this type have relied on much more restrictive structural assumptions.

Our results directly extend and unify previous results on intermittent observations and 
sensor scheduling, and characterize the stabilizability of systems for varying intermittency rates. 
We show this extension explicitly, and also provide an explicit definition of the stabilizable 
set of intermittency rates for a simple but useful example.

The geometric convergence and other structural characteristics of the control results
also suggest more general analytical results should be achievable, and we anticipate
extensions of the structural results for specific distributed control problems.



\end{document}